\documentclass[prx,superscriptaddress,reprint]{revtex4-2}

\usepackage{amsmath}
\usepackage{graphicx}
\usepackage{bm}
\usepackage{MnSymbol}
\usepackage{eufrak}
\usepackage{amsfonts} 

\usepackage[usenames,dvipsnames,x11names]{xcolor}
\usepackage[normalem]{ulem}
\usepackage{tikz}
\usetikzlibrary{arrows}
\usepackage{mathrsfs}
 \usepackage{color}
 \definecolor{darkblue}{rgb}{0,0,.5}
 \usepackage[linktocpage, colorlinks=true, linkcolor=darkblue, citecolor=darkblue]{hyperref}
 \usepackage[all]{hypcap}
 \usepackage[makeroom]{cancel}

\newcommand{\Dalpha}{D[\boldsymbol \alpha(t)]}
\newcommand{\Ddalpha}{D^\dagger[\boldsymbol \alpha(t)]}
\newcommand{\ket}[1]{\left\vert #1    \right\rangle }
\newcommand{\bra}[1]{\left\langle   #1  \right\vert}
\newcommand{\braket}[1]{\left\langle   #1  \right\rangle}
\newcommand{\new}[1]{{ {#1}}}

\begin{document}

\title{Topological synchronization of quantum van der Pol oscillators}

\author{Christopher W. W\"achtler}
\email{cwwaechtler@berkeley.edu}
\affiliation{Max Planck Institut für Physik komplexer Systeme, 01187 Dresden, Germany}
\affiliation{Department of Physics, University of California, Berkeley, California 94720, USA}
\author{Gloria Platero}
\address{Instituto de Ciencia de Materiales de Madrid, CSIC, Madrid 28049, Spain}

\date{\today}

\begin{abstract}
To observe synchronization in large networks of classical or quantum systems demands both excellent control of the interactions between nodes and very accurate preparation of initial conditions due to the involved nonlinearities and dissipation. This limits its applicability for future devices. We demonstrate a route towards significantly enhancing the robustness of synchronized behavior in open nonlinear systems that utilizes the power of topology. In \new{non-trivial topological} lattices of quantum van der Pol oscillators, boundary synchronization emerges in the classical mean field as well as the quantum regime. In addition to its robustness against disorder and initial state perturbations, the observed dynamics is independent of the underlying topological model provided the existence of topological zero-energy modes. Our work extends the notion of topology to the general nonlinear dynamics and open quantum system realm with applications to networks where specific nodes need special protection like power grids or quantum networks. 
\end{abstract}

\maketitle

For many quantum mechanical applications dissipation is often regarded as an undesirable yet unavoidable consequence because it potentially degrades quantum coherences and renders the system classical. However, interactions with the environment can also be considered a fundamental resource for striking collective effects typically impossible in Hamiltonian systems \cite{PhysRevA.78.042307, PhysRevX.8.021005, PhysRevLett.123.030603}. A hallmark of such collective behavior in nonequilibrium systems is the phenomenon of synchronization: in the complete absence of any time-dependent forcing from the outside, a group of oscillators adjusts their frequencies such that they spontaneously oscillate in unison \cite{pikovsky2003synchronization,StrogatzBook2018}. Synchronization is intimately related to the phenomenon of self-sustained oscillations, where a system maintains a periodic motion in an autonomous fashion \cite{JenkinsPR2013}. In this sense, interactions with the environment and nonlinearities in the equations of motion represent prerequisites for synchronization. With the recent developments in quantum technology which allow one to exquisitely tailor both the system and environmental properties, synchronization has emerged in the quantum domain with various different examples ranging from nonlinear oscillators to spin-1 systems, superconducting qubits, optomechanics and ensembles of atoms that demonstrate synchronization in the quantum regime \cite{Lee2013,Lee2014,PhysRevLett.112.094102,Walter2015,bastidas2015,PhysRevLett.113.154101,PhysRevLett.121.063601,sonar2018squeezing,PhysRevLett.121.053601,es2020synchronization,cattaneo2021bath,PhysRevLett.129.063605, tan2022half}. 

In both quantum and classical systems, unavoidable imperfections – local deformations caused by ambient conditions as well as long-term degradation – have a significant impact on the collective behavior and can even destroy synchronicity altogether. Moreover, collective synchronized dynamics in large networks often sensitively depends on fine-tuned initial conditions. It is therefore desirable for a reliable performance of future devices to identify universal principles to enhance the robustness of synchronization. In this work, we demonstrate that the power of topology can be exploited for this task. Topological insulators describe a special class of solids exhibiting an insulating bulk but symmetry protected conducting surface states, known as topological edge states \cite{thouless1982quantized,haldane1988model,moore2010birth,HasanKaneRMP2010,hasan2011three}. These edge states display a surprising immunity to a wide range of local deformations, inherently avoiding backscattering over broad energy ranges and even circumventing localization in the presence of disorder \cite{MittalEtAlPRL2014}. 

While the robustness of topological systems is well established for isolated systems, it is a nontrivial problem to determine whether this feature is still available in open setups. Recently, the study of out-of-equilibrium systems has moved to the forefront of research in topological band structures, motivated largely by the desire to create interesting states of matter with properties beyond those achievable in equilibrium \cite{Diehl2011,Bardyn2013,PhysRevX.8.031079,PhysRevLett.121.026808,PhysRevLett.121.086803,PhysRevX.9.041015,song2019non}. While topological band  and bandgap structures are inherently tied to linear systems, we currently observe rapidly growing interest to generalize topological concepts to nonlinear systems, which in turn has sparked the field of nonlinear topological photonics and the phenomenon of topological lasing \cite{smirnova2020nonlinear,PhysRevLett.118.023901,Harari2018,Bandres2018,wang2019topologically}.

\new{Despite these recent advancements, applying topological concepts to synchronization phenomena is still largely unexplored, with only a few studies on classical nonlinear oscillators \cite{WaechtlerEtAlPRB2020,kotwal2021active,sone2022topological}, yet with already some remarkable results. For example, Sone et al. \cite{sone2022topological} showed that linearly coupled Stuart-Landau oscillators arranged on a square lattice may exhibit chaotic bulk motion and robust synchronized edge oscillations when the coupling Hamiltonian promotes topological lasing modes. The authors termed this coexistence state `topological synchronized state', which may be observed for Hermitian as well as non-Hermitian coupling Hamiltonians. Furthermore, the involved nonlinearity may also induce emerging effective boundaries, which lead to extra topological modes unique to nonlinear systems. However, to the best of our knowledge, similar studies for quantum systems are entirely missing so far. In this work, we systematically investigate the potential of topological insulator models to promote \textit{topological synchronization} in classical and quantum systems.} In order to provide such a comprehensive study, we choose the van der Pol (vdP) oscillator \cite{StrogatzBook2018,pikovsky2003synchronization} as our building block to construct topological lattices. Since the quantum version of the vdP oscillator \cite{Lee2013,Lee2014,Walter2015} reduces to its classical analog  at the mean field level (specifically to the Stuart-Landau oscillator, which is the weakly nonlinear limit of the classical vdP oscillator), we are able to study the influence of topological band structures in both regimes. We show not only that boundary synchronization emerges in the classical as well as the quantum regime if the underlying lattice possesses nontrivial topology, but also that it is protected over a wide range of local disorder. Our results pave the way towards utilizing topology as a powerful tool for enhancing the robustness of collective dynamics in nonlinear open quantum systems.

\section{Topological synchronization in coupled oscillator networks}
\label{sec:Theory}

\subsection{Topological van der Pol oscillator network}
\label{subsec:FullModel}

We study a lattice of $N$ sites, each consisting of a harmonic oscillator labeled by the index $j=1,...,N$ and we assume identical frequencies $\omega_0$ of each oscillator. The Hamiltonian of the system can be written as a tight-binding Hamiltonian 
\begin{equation}
\label{eq:GeneralHamiltonian}
H=H_0 + H_\mathrm{top} = \hbar\sum\limits_j \omega_0 a_j^\dagger a_j + \hbar\sum\limits_j \sum_{j'\neq j} \lambda_{jj'}(a_j^\dagger a_{j'} + a_{j'}^\dagger a_j) ,
\end{equation} 
where $a_j^\dagger$ ($a_j$) denote creation (annihilation) operators of bosonic particles at lattice site $j$. This general form of the system Hamiltonian allows us to realize different topological lattices simply by modifying the couplings $\lambda_{jj'}$. Recently, it has been realized that a quantum harmonic oscillator subject to one-phonon gain with rate $\kappa_1$ and two-phonon loss with rate $\kappa_2$ represents the quantum analogue of the classical vdP oscillator \cite{Lee2013,Lee2014,PhysRevLett.112.094102,Walter2015,bastidas2015}. In an open quantum system approach and using the notation $\mathcal D\left[O\right]\varrho=O \varrho O^\dagger - \frac{1}{2}\left\{O^\dagger O, \varrho\right\}$, the dynamics of the system density matrix $\varrho(t)$ describing a network of vdP oscillators is then given by the master equation 
\begin{equation}
\label{eq:QauntumModel}
\dot{\varrho} = -\frac{\mathrm i}{\hbar}\left[H,\varrho\right] +  \sum\limits_{j} \left\{\kappa_1\mathcal D\left[a_j^\dagger\right]\varrho +\kappa_2\mathcal D\left[a_j^2\right]\varrho \right\}.
\end{equation}

Throughout this work we focus on the weak dissipation regime, that is we assume that $\kappa_1, \kappa_2 \ll \omega_0$. For realistic systems, this regime corresponds to long coherence times and weak coupling to the reservoirs such that a description of the system dynamics in terms of a Lindblad master equation [cf. Eq.~(\ref{eq:QauntumModel})] is valid. Furthermore, we are interested in the regime where $\kappa_2> \kappa_1$ as in this case each oscillator remains close to its ground state \cite{Lee2013}. 

\section{Methods}
\subsection{Mean field approximation and linear stability analysis}
\label{subsec:MeanField}
From the full quantum model defined in Eq.~(\ref{eq:QauntumModel}), one may derive classical equations of motion for the expectation values $\alpha_j=\langle a_j\rangle\in\mathbb C$ by performing a mean field approximation. The governing equation of the complex-valued mean field amplitudes $\boldsymbol \alpha=(\alpha_1,\dots,\alpha_N)$ is then given by 
\begin{equation}
\label{eq:VdPLattice}
\dot{\boldsymbol \alpha} = -\frac{\mathrm i}{\hbar} \underline{\mathrm{H}}\boldsymbol \alpha + \frac{\kappa_1}{2} \boldsymbol \alpha - \kappa_2 \left(\boldsymbol \alpha \odot \boldsymbol \alpha^\ast \odot\boldsymbol \alpha \right),
\end{equation}
where $\underline{\mathrm{H}}= \underline{\mathrm H}_0+\underline{\mathrm{H}}_\mathrm{top}$ denotes the matrix corresponding to the Hamiltonian $H$ [cf. Eq.~(\ref{eq:GeneralHamiltonian})] with diagonal matrix $\underline{\mathrm H}_0 = \hbar \omega_0 \underline{1}$ and  the matrix describing the coupling between oscillators $\underline{\mathrm{H}}_\mathrm{top}$. Furthermore, $\odot$ denotes the Hadamard product defined as \new{$(\boldsymbol \alpha\odot \boldsymbol \alpha)_{n} = \boldsymbol \alpha_{n}\cdot \boldsymbol \alpha_{n}$}. 

In the absence of any coupling between lattice sites ($\lambda_{j{j'}}\equiv 0$) each oscillator will approach its respective periodic steady state $\bar \alpha_j(t) = \bar A \exp(-\mathrm i \omega_0 t+\varphi_j)$ with $\bar A =\sqrt{\kappa_1/2\kappa_2}$ and arbitrary phases $\varphi_j$. However, the interactions enable the emergence of collective synchronized motion, whose existence \new{is intimately related to the stability of the fixed point $\boldsymbol \alpha_\mathrm{FP}$ defined via $\dot{\boldsymbol \alpha}_\mathrm{FP}=0$. From Eq.~(\ref{eq:VdPLattice}), we find that the fixed point is given by $\boldsymbol \alpha_\mathrm{FP}=0$. The stability of $\boldsymbol \alpha_\mathrm{FP}$ may be analyzed in terms of a corresponding model linearized around $\boldsymbol \alpha_\mathrm{FP}$: }
\begin{equation}
\label{eq:Linearized}
\new{\dot {\boldsymbol \alpha} = \underline{\mathrm{J}} \boldsymbol \alpha,} 
\end{equation}
\new{where we have omitted second and higher order terms.}
The Jacobian matrix $\underline{\mathrm J}$ with entries $\underline{\mathrm J}_{jj'}= \partial \dot\alpha_j /\partial \alpha_{j'}$ is evaluated at ${\boldsymbol \alpha}_\mathrm{FP}$ resulting in the matrix
\begin{equation}
\underline{\mathrm{J}} =  -\frac{\mathrm i}{\hbar}\left(\underline{\mathrm H}_0+\underline{\mathrm{H}}_\mathrm{top}\right) +\frac{\kappa_1}{2}\underline{1}.
\end{equation} 
\new{In particular, Eq.~(\ref{eq:Linearized}) only contains the linear terms of Eq.~(\ref{eq:VdPLattice}). In general, the fixed point $\boldsymbol \alpha_\mathrm{FP}$ is stable as long as the real part of all eigenvalues of $\underline{\mathrm{J}}$ are negative. In contrast, if at least one eigenvalue has positive real part, the fixed point is dynamically unstable and the corresponding eigenstate will exponentially grow. However, the nonlinear damping in Eq.~(\ref{eq:VdPLattice}) counteracts this exponential amplitude increase, leading eventually to stable oscillations.}

Since $\underline{\mathrm{H}}_\mathrm{top}$ is a Hermitian matrix, it has real eigenvalues $\hbar \mu^{(l)}$, where $l=1,\dots, N$ is an index for the different eigenvalues. Moreover, the diagonal entries of $\underline{\mathrm{J}}$ are given by $-\mathrm i \omega_0+\kappa_1/2$, such that $\nu^{(l)} = -\mathrm i (\omega_0+\mu^{(l)}) + \kappa_1/2$ are the eigenvalues of $\underline{\mathrm{J}}$. \new{Hence, for $\kappa_1>0$ (which we assume throughout this work), all eigenstates are linearly unstable. Moreover,} for $\kappa_2 > \kappa_1$ we expect small oscillation amplitudes which remain close to the fixed point even if the coupling $\lambda_{j{j'}}$ is comparable to the intrinsic frequency $\omega_0$. Consequently, the dynamics  of the oscillator lattice (after relaxation) \new{may be} well approximated  by a superposition of eigenvectors,
\begin{equation}
\label{eq:DynLS}
\boldsymbol \alpha_\mathrm{lin}(t) = \sum\limits_ l c^{(l)} \boldsymbol \alpha^{(l)} e^{-\mathrm i [\omega_0 +\mu^{(l)}] t}, 
\end{equation}
where $\mu^{(l)}$ and $\boldsymbol \alpha^{(l)}$ are the eigenvalues and their respective eigenvectors of $\underline{\mathrm{H}}_\mathrm{top}/\hbar$, and $c^{(l)}$ denotes scalar coefficients. \new{This may suggest that the system dynamics is fully described in terms of the linear model and thus by Eq.~(\ref{eq:Linearized}). However, our results -- discussed in Sec.~\ref{sec:Results} -- indicate that such a simple picture is not sufficient to capture the full dynamical evolution of the system in the presence of dissipation and nonlinearities.}

\subsection{Quantum fluctuations}
\label{subsec:EffectiveModel}
\new{Solving the full open quantum system given by Eqs.~(\ref{eq:GeneralHamiltonian}) and (\ref{eq:QauntumModel}) is a non-trivial task due to the large number of interacting oscillators and the involved nonlinearities in the dissipators. In order to gain further insight, we follow a similar derivation to the ones   found in Refs.~\cite{PhysRevX.4.011015,bastidas2015}: We move} to a displaced frame via the displacement operator 
\begin{equation}
D[\boldsymbol \alpha (t)] = \exp \left\{\sum\limits_j \left[\alpha_j(t) a_j - \alpha_j^\ast(t) a_j^\dagger \right]\right\}
\end{equation}
and \new{define} the density matrix in the displaced frame as $\varrho_\alpha(t) = D^\dagger[\boldsymbol \alpha(t)]\varrho(t) D[\boldsymbol \alpha(t)]$. \new{Interestingly, the terms linear in the operators $a_j$ and $a_j^\dagger$ vanish as long as Eq.~(\ref{eq:VdPLattice}) is fulfilled, such that we are left with terms quadratic and cubic in the operators; see App.~\ref{secApp:EffectiveQM} for details.} By neglecting terms of order $\mathcal O(a_j^3)$, we obtain an effective master equation of Lindblad form, 
\new{\begin{equation}
\label{eq:MasterEquationAlpha}
\begin{aligned}
\dot \varrho_\alpha(t) =& -\frac{\mathrm i}{\hbar} \left[H_\alpha(t), \varrho_\alpha(t) \right] \\
&+\sum\limits_j \left\{\kappa_1 \mathcal D\left[ a_j^\dagger\right]\varrho_\alpha(t)+ 4 \kappa_2 \left|\alpha_j(t)\right|^2\mathcal D\left[ a_j\right]\varrho_\alpha(t)\right\}
\end{aligned}
\end{equation}}
with effective\new{, time-dependent} (squeezing) Hamiltonian 
\new{\begin{equation}
\label{eq:EffectiveHamiltonian}
H_\alpha(t) = H - \mathrm i\hbar\frac{\kappa_2}{2}\sum\limits_j \left\{\left[\alpha_j(t)\right]^2  (a_j^\dagger)^2 - \left[\alpha_j^\ast(t)\right]^2  a_j^2\right\}.
\end{equation}}
\new{Note, that the full dynamics of the system is now determined by Eq.~(\ref{eq:MasterEquationAlpha}) in combination with Eq.~(\ref{eq:VdPLattice}) because the time-dependent mean field amplitudes $\alpha_j(t)$ appear in the effective Hamiltonian (\ref{eq:EffectiveHamiltonian}) and the time-dependent dissipation rates. Hence, in general both equations have to be solved simultaneously. However, as Eq.~(\ref{eq:VdPLattice}) is independent of the dynamics of the density matrix $\varrho_\alpha(t)$, the mean field dynamics may be solved first and treated as time-dependent inputs for Eq.~(\ref{eq:MasterEquationAlpha}). In App.~\ref{secApp:Validity}, we compare this effective model to the full quantum model described by Eq.~(\ref{eq:QauntumModel}) for a single and two coupled quantum vdP oscillators in terms of Wigner functions.}

As the Hamiltonian~(\ref{eq:EffectiveHamiltonian}) is quadratic, the dynamics of the effective quantum model is fully described by the covariance matrix \new{$\underline{\mathrm C}_{mn}(t)=\mathrm{Tr[\varrho_\alpha(t) \{X_m, X_{n}\}/2]}$} with the quadratures $X_{2j-1}=( a_j+ a^\dagger_j)/\sqrt{2}$ and $X_{2j}=-\mathrm i ( a_j -  a_j^\dagger)/\sqrt{2}$. The equation of motion of the covariance matrix \new{$\underline{\mathrm C}(t)$} is given by \cite{PhysRevA.95.041802}
\new{\begin{equation}
\label{eq:DiffEqCorrelation}
\dot{\underline{\mathrm C}}(t) = \underline{\mathrm B}(t)~\underline{\mathrm C}(t)+ \underline{\mathrm C}(t)~\underline{\mathrm B}^\intercal(t) + \underline{\mathrm D}(t), 
\end{equation}}
where \new{$\underline{\mathrm B}(t)$} and \new{$\underline{\mathrm D}(t)$} are determined from Eq.~(\ref{eq:MasterEquationAlpha}); see App.~\ref{secApp:Covariance}. 
Throughout this work, we consider a pure coherent state as the initial state $\varrho(0) = \bigotimes_j \ket{\alpha_j(0)}\bra{\alpha_j(0)}$.  In the comoving frame, such an initial condition corresponds to $\varrho_\alpha(0) = \bigotimes_j \ket{0_j}\bra{0_j}$ which results in an initial covariance matrix of diagonal form, i.e., $\underline{\mathrm C}(0) = 1/2\cdot \underline{1}$, which reflects the Heisenberg uncertainty principle. \new{Note, that the time-dependency of $\underline{\mathrm B}(t)$ and $\underline{\mathrm D}(t)$ arises from the time-dependent mean field amplitudes $\alpha_j(t)$ (see also see App.~\ref{secApp:Covariance}), such that Eqs.~(\ref{eq:VdPLattice}) and (\ref{eq:DiffEqCorrelation}) are also solved simultaneously as mentioned previously.}

\subsection{Synchronization in classical and quantum systems}
\label{subsec:IntoSynch}
\new{Synchronization of classical oscillatory systems can take many different forms. A primary example often discussed is the case of two strongly nonlinear systems which oscillate with different natural frequencies, and that adjust their rhythm due to a mutual weak coupling referred to as frequency synchronization. Additionally, the two oscillators may also synchronize their phases such they oscillate in-phase (or anti-phase). In the case of weak coupling, one can show that only phases rather than amplitudes of oscillators are relevant (see e.g. \cite{kuramoto1984chemical}). Conversely, a relatively strong coupling between two oscillators can affect not only their phases but also their amplitudes, and the features of synchronization in the presence of strong interactions are nonuniversal \cite{pikovsky2003synchronization}. For example, on a finite lattice even if all oscillators are identical and they only interact locally with their nearest neighbors, synchronicity across the lattice is not guaranteed; examples will be discussed later in Sec.~\ref{sec:Results}. }

\new{In this work, we are concerned to what extent topological lattices and, in particular, edge states affect the synchronicity of self-oscillatory systems. As the lattices studied here host edge states that are zero dimensional, the corresponding oscillators are located at the boundaries and only interact via many bulk oscillators. If these bulk oscillators are neither frequency nor phase synchronized, one may not expect that the oscillators are able to synchronize their phases. However, they might still be able to oscillate with a common frequency. Thus, throughout this work we will refer to two classical systems $j$ and $j'$ being synchronized if the condition 
\begin{equation}
\label{eq:SynchPhiCondition}
\frac{d}{dt}\Delta \varphi_{jj'} = \frac{d}{dt}\left(\varphi_j -\varphi_{j'}\right)=0,
\end{equation}
is fulfilled. Note, that if this conditions is fulfilled for all $j$ and $j'$, the whole lattice oscillates with a single common frequency which is expected to be an eigenfrequency of the lattice (i.e. $\omega_0 + \mu^{(l)}$; see Sec.~\ref{subsec:MeanField}) because of the weakly nonlinear regime.} 

Generalizing the previously introduced notion of synchronization to the quantum regime is challenging as phase space trajectories become ill defined concepts. To overcome this challenge, synchronization measures in terms of the Husimi-Q or Wigner phase space distributions \cite{Lee2013,PhysRevLett.121.063601,parra2020synchronization}, explicit limit cycles of system observables \cite{Tindall2020,buca2022algebraic}, or information-theoretical measures \cite{ameri2015mutual,jaseem2020generalized} have been proposed. In this work, we use a measure to quantify synchronization of two quantum systems $j$ and $j'$ that is based on their dimensionless quadratures as \cite{2013-Fazio-PRL}
\new{\begin{equation}
\label{eq:CompleteSynch}
S_\mathrm c(j,j') = \left<(X_{2j-1}-X_{2j'-1})^2+(X_{2j}-X_{2j'})^2\right>^{-1}\leq 1, 
\end{equation}}
where $X_{2j-1}=( a_j+ a^\dagger_j)/\sqrt{2}$ and $X_{2j}=-\mathrm i ( a_j -  a_j^\dagger)/\sqrt{2}$. The upper bound of the quantum synchronization measure arises from the Heisenberg uncertainty principle. \new{The synchronization measure $S_\mathrm c(j,j')$, also referred to as `complete synchronization measure' \cite{2013-Fazio-PRL}, extends the classical concept of `error' into the quantum domain, i.e., the smaller the variance of the quadrature differences is, the larger is the synchronization measure $S_\mathrm c(j,j')$. It thus indicates how equivalent the dynamics of two quantum systems are, which in turn connects to the classical notion of two systems oscillating with the same frequency in-phase or anti-phase. Here, $S_\mathrm c(j,j')$ would take on its maximum value; see also App.~\ref{secApp:Validity}. }

As $S_\mathrm c(j,j')$ may exhibit oscillations even in the long time limit, we perform time averages of this quantity, i.e.,
\begin{equation}
\label{eq:AvgCompleteSynch}
\left< S_\mathrm{c}(j,j')\right> = \frac{1}{t_\mathrm f- t_\mathrm i}\int\limits_{t_\mathrm i}^{t_\mathrm{f}} S_\mathrm c(j,j') dt'.
\end{equation}
with initial time $t_i$ and final time $t_f$. Throughout this work we are interested in the (periodic) steady state dynamics and discard transient dynamics, such that both $t_i$ and $t_f$ are taken to be larger than the transient relaxation time (see Sec.~\ref{subsec:Prelim}).

\section{Results}
\label{sec:Results}

\subsection{Preliminary remarks}
\label{subsec:Prelim}
Before we turn to the detailed analysis of two specific topological lattices in the next two sections, we would like to provide some general notions that will apply to both examples in order to minimize redundancies. 
Within each section, $\boldsymbol \alpha^{(l)}$ indicates an eigenstate with eigenvalue $\hbar\mu^{(l)}$ of the respective topological (coupling) Hamiltonian, i.e., $H_\mathrm{SSH}$ in Example (I) and $H_\mathrm{Kag}$ in Example (II). Furthermore, the mean field amplitude of lattice site $j$ is denoted by $A_j(t)$ and defined via $A_j(t) = [\alpha_j(t) + \alpha_j^\ast(t)]/2$\new{, which is proportional to the averaged position of each oscillator}. Throughout, when random initial conditions are considered, we mean randomly distributed complex amplitudes and phases, i.e., $a_j(t=0) = \alpha_j^{\mathrm r} \exp{(i\varphi_j^\mathrm r)}$ where $\alpha_j^{\mathrm r}$ and $\varphi_j^\mathrm r$ are random variables chosen from the uniform distributions $\alpha_j^{\mathrm r}\sim \mathcal U(0,0.5)$ and $\varphi_j^\mathrm r\sim \mathcal U(0,2\pi)$, respectively. Similarly, when random disorder with strength $r$ is applied to the coupling between lattice sites $j$ and $j'$ we refer to \new{$\lambda_{jj'} \to \lambda_{jj'} + \delta \lambda_{jj'}^\mathrm r$} with $ \delta \lambda_{jj'}^\mathrm r\sim \mathcal U(-r,r)$. Lastly, all numerical results shown are obtained after a significant relaxation time $\omega_0 t_\mathrm{rel} = 2\cdot 10^{4}$ in order to discard transient effects. 

\subsection{Example (I): 1D SSH chain}
\label{sec:SSH}
\subsubsection{The SSH model}
\label{subsec:SSHIntro}

\begin{figure*}
\begin{center}
\includegraphics[width=\linewidth]{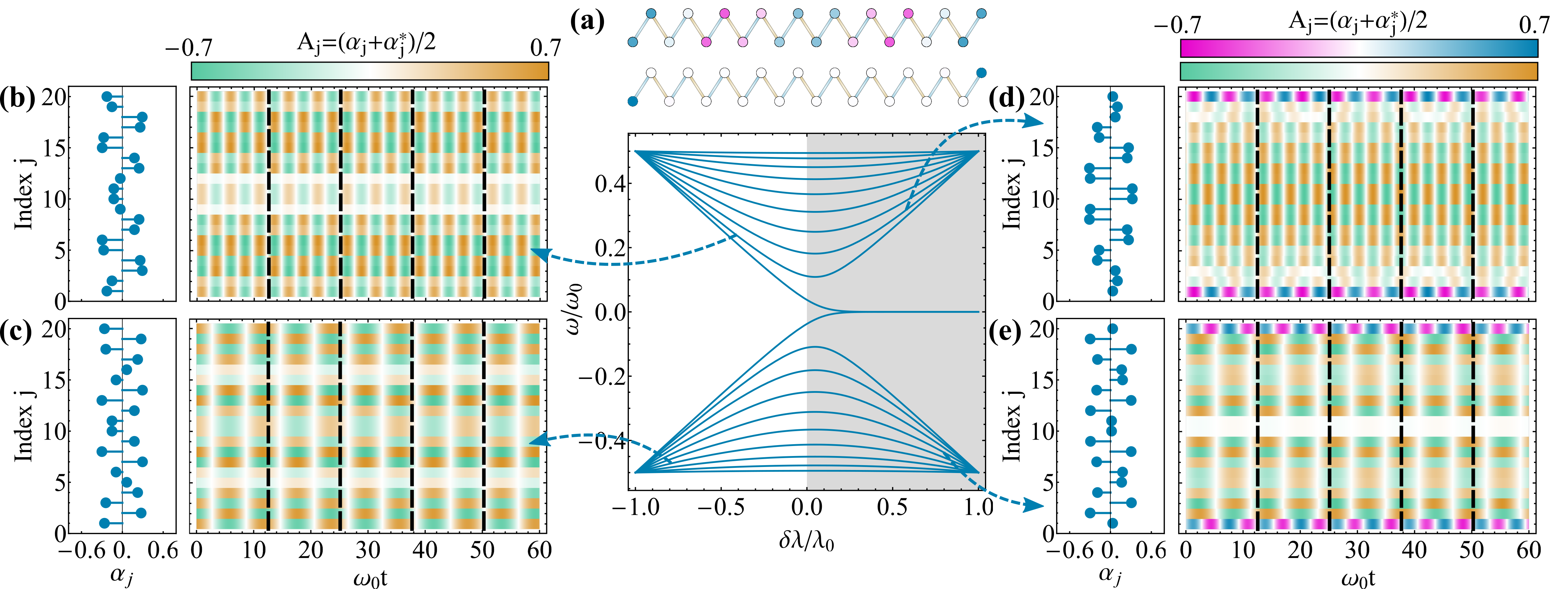}
\end{center}
\caption{(a) Eigenspectrum of $H_\mathrm{SSH}$ (or equivalently $\underline{\mathrm H}_\mathrm{SSH}$) for $N=20$ with $\lambda _1=\lambda _0-\delta \lambda $ and $\lambda _2=\lambda _0+\delta \lambda $, where the gray region marks the topological phase ($\lambda _2>\lambda _1$). The top two lattices show examples of a  delocalized bulk state (top) and a highly localized edge state (bottom) for $\delta \lambda /\lambda_0 = 0.8$.  (b)--(e) Left panels show the amplitudes $\alpha_j^{(l)}$ of the eigenvectors corresponding to the eigenvalues $\hbar \mu^{(l)}$ marked by the blue arrows of the eigenspectrum (a), specifically (b) $\mu^{(l)}=0.94 \lambda_0$ for $\delta \lambda /\lambda _0 = -0.4$, (c) $\mu^{(l)}=-1.89 \lambda _0$ for $\delta \lambda /\lambda _0 = -0.8$, (d) $\mu^{(l)}=1.23 \lambda _0$ for $\delta \lambda /\lambda _0 = 0.6$, and (e) $\mu^{(l)}=-1.87 \lambda _0$ for $\delta \lambda /\lambda _0 = 0.8$. These eigenstates are the initial states for the mean field dynamics of the vdP network. The amplitude dynamics $A_j(t)$  of each oscillator $j$ is shown in the right panels \new{after the relaxation time $\omega_0 t_\mathrm{rel}$}. In the nontrivial topological phase [panels (d) and (e)] the two oscillators located at the edges ($j=1$ and $j=20$) are highlighted in blue/pink. Parameters: $\kappa_1 = 5\cdot 10^{-3}\omega_0$, $\kappa_2 = 2\kappa_1$, $\lambda _0 = 0.25\omega_0$.}
\label{fig:Fig2}
\end{figure*}

In the following we investigate the interplay of topology and synchronization in the paradigmatic SSH model \cite{HeegerEtAlRMP1988,topologicalInsulators}, a one-dimensional dimerized lattice with staggered nearest-neighbor hopping and time reversal, particle-hole and chiral symmetry (BDI class). Recently, implementations of this model under nonequilibrium conditions or in strongly nonlinear systems have been proposed \cite{Gomez-Leon2013,engelhardt2016topological,bello2016long,engelhardt2017topologically,bello2019unconventional} and experimentally realized \cite{PhysRevB.93.155112, hadad2017solitons, PhysRevLett.121.163901, hadad2018self, PhysRevA.100.063830, kim2021quantum}. The Hamiltonian of the model is given by 
\begin{equation}
\label{eq:SSHHamiltonian}
H_\mathrm{SSH} = \hbar\sum_j \lambda_j (a_j^\dagger a_{j+1} + a_{j+1}^\dagger a_j),
\end{equation}
where $\lambda_j=\lambda_1$ if $j$ is odd and $\lambda_j = \lambda_2$ otherwise. The spectrum of the Hamiltonian~(\ref{eq:SSHHamiltonian}) is symmetric with respect to the zero-energy axis ($E=0$) as a result of the chiral symmetry present in the system. A phase transition of topological nature separates the trivial phase ($\lambda_1>\lambda_2$) from the topological phase ($\lambda_1<\lambda_2$). In Fig.~\ref{fig:Fig2}(a) we show the eigenspectrum of $H_\mathrm{SSH}$ for $N=20$ lattice sites as a function of $\delta \lambda$, where $\lambda_1 =\lambda_0 -\delta \lambda$ and $\lambda_2 = \lambda_0+\delta \lambda$. For the chain with finite length, eigenvectors within the bands are delocalized along the whole chain; examples are shown in Fig.~\ref{fig:Fig2}(a) and in the left panels of Figs.~\ref{fig:Fig2}(b)--(e). 

In addition, two degenerate zero-energy states occur within the band gap in the topological phase ($\delta \lambda >0$), which are highly localized at the boundaries of the system [Fig.~\ref{fig:Fig2}(a)]. The edge states are topologically protected by the chiral symmetry and are thus robust against (symmetry-preserving) local perturbations as we discuss in more detail below. 

\subsubsection{Topological effects on synchronization at the mean field level}
\label{subsec:TopologySSH}

To incorporate the topological character of the SSH model into the network of vdP oscillators, the coupling matrix $\underline{\mathrm H}_\mathrm{top}$ in Eq.~(\ref{eq:VdPLattice}) is chosen to be  $\underline{\mathrm H}_\mathrm{SSH}$. In the following we compare the numerical results to our previous analytic considerations. In the right panels of Figs.~\ref{fig:Fig2}(b)--(e) we show the amplitude dynamics $A_j(t)$ of the oscillator at site $j$ when an eigenstate $\boldsymbol \alpha^{(l)}$ of $\underline{\mathrm H}_\mathrm{SSH}$ is chosen as initial state, shown in the left panels of Figs.~\ref{fig:Fig2}(b)--(e). 

We start our discussion in the trivial phase ($\delta \lambda<0$). In Figs.~\ref{fig:Fig2}(b) and (c), we observe complete synchronization, i.e. all lattice sites $j$ oscillate with the same frequency $\omega_0 + \mu^{(l)}$. This frequency is determined by the choice of initial eigenstate $\boldsymbol \alpha^{(l)}$ with respective eigenvalue $\hbar \mu^{(l)}$. Hence, the oscillation frequency in panel (b) [(c)] is larger (smaller) compared to the intrinsic frequency $\omega_0$ of the uncoupled vdP oscillators as $\mu^{(l)} >0$ ($<0$). Furthermore, the amplitude of the initial state $\alpha_{j}^{(l)}$ directly translates to the oscillation amplitudes $A_j(t)$ and their phases. Thus in accordance to our previous analytic considerations, the dynamics in the trivial phase  is given by $\boldsymbol \alpha(t) = c \boldsymbol \alpha^{(l)} \exp\{-\mathrm i [\omega_0 + \mu^{(l)}]t\}$ with a scaling factor $c$. 

 \begin{figure*}
\begin{center}
\includegraphics[width=\linewidth]{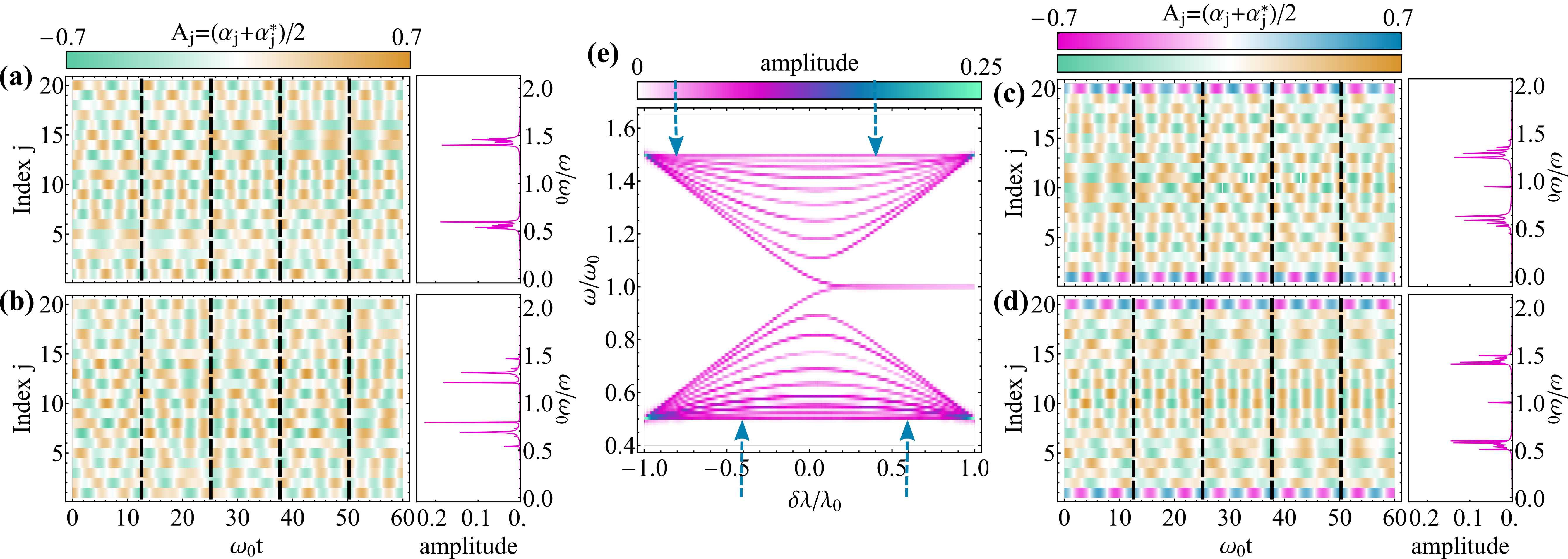}
\end{center}
\caption{(a)--(d) Left panels show the amplitude dynamics $A_j(t)$  of each oscillator $j$ of the vdP network \new{after the relaxation time $\omega_0 t_\mathrm{rel}$} for random initial conditions for different couplings (a) $\delta \lambda=-0.8 \lambda_0$, (b) $\delta \lambda = - 0.4\lambda_0$, (c) $\delta \lambda = 0.4 \lambda_0$, and (d) $\delta \lambda = 0.6 \lambda_0$ [also marked with blue arrows in panel (e)]. The two oscillators located at the edges ($j=1$ and $j=20$) are highlighted in blue/pink in panels (c) and (d). Right panels show the frequency spectrum obtained from a discrete Fourier transform of the dynamics of the left panels. (e) Reconstruction of the eigenspectrum of \new{$H_0+H_\mathrm{SSH}$} (or equivalently \new{$\underline{\mathrm H}_0+\underline{\mathrm H}_\mathrm{SSH}$}) from the oscillation dynamics averaged over 10 realizations of random initial conditions.  Parameters: $\kappa_1 = 5\cdot 10^{-3}\omega_0$, $\kappa_2 = 2\kappa_1$, $\lambda _0 = 0.25\omega_0$.}
\label{fig:FigFourier}
\end{figure*}

For $\delta \lambda> 0$, however, nontrivial topological effects on the dynamics of the lattice emerge as shown in Figs.~\ref{fig:Fig2}(d) and (e): Even though the initial bulk eigenstate (shown in the left panels) exhibits only small amplitudes at the edges ($j=1$ and $j=20$) their oscillation amplitudes (highlighted in blue/pink) are comparable to the largest amplitudes in the whole chain. Moreover, they  oscillate with the intrinsic frequency $\omega_0$ (the vertical dashed lines are located at integer multiples of $\omega_0 t = 4\pi n$, $n\in\mathbb N$). In comparison to the trivial phase, where small initial amplitudes remain small for all lattice site, the edge modes in the topological phase are always excited. Note that the edge states are not strictly localized at the two boundary lattice sites, but rather extend exponentially into the bulk. Consequently, the dynamics of bulk oscillators close to the edge, e.g. $j=3$ or $j=18$, is a superposition of the initial state and the edge state with different frequencies. Thus, the dynamics of these oscillators is given by $\alpha_j(t) \propto \alpha_j^{(l)} \exp\{-\mathrm i[\omega_0 + \mu^{(l)}]t\} + \alpha_j^{\mathrm edge} \exp(-\mathrm i\omega_0 t)$, where $\alpha_j^{(l)}$ and $\alpha_j^{\mathrm edge}$ represent the initial and edge eigenstate, respectively. 

Since it might be difficult in an experimental setup to initiate the system in a specific eigenstate $\boldsymbol \alpha^{(l)}$ of $ \underline{\mathrm H}_\mathrm{SSH}$, the question arises how a randomly distributed initial state affects the dynamics and whether the previously discussed zero-energy edge mode synchronization persists in such a scenario. To this end, we show in  the left panels of Figs.~\ref{fig:FigFourier}(a)--(d) the amplitude dynamics $A_j(t)$ for such random initial states. In the trivial phase shown in panels (a) and (b) there is no clear pattern of synchronization. We would like to note that some oscillators in Figs.~\ref{fig:FigFourier}(a) and (b) might be synchronized for short times, however, their phase difference is not constant over longer times. 

In order to confirm our intuition that there are many frequencies participating in the dynamics of the vdP network, we perform a discrete Fourier transform and show the frequency spectrum in the right panels of Figs.~\ref{fig:FigFourier}(a) and (b). As expected there are multiple peaks centered around $\omega_0$ and separated by a gap. These peaks correspond to the eigenfrequencies $\omega_0 +\mu^{(l)}$ of the system. 

In the topological phase shown in Figs.~\ref{fig:FigFourier}(c) and (d) the oscillator dynamics (left panels) is similar to the one described previously for the trivial phase. However, there is an important difference: The edges (highlighted in blue/pink) do not exhibit frequency mixing but oscillate with one specific frequency, the intrinsic frequency $\omega_0$. Hence, the edges are synchronized with one another, yet due to the random initial conditions with a phase shift difference (e.g. different amplitudes at the dashed lines for $j=1$ and $j=20$). However, this phase difference remains constant over time. In the frequency spectrum of the discrete Fourier analysis, we observe now an additional sharp peak at $\omega=\omega_0$ confirming the existence of the synchronized edge modes. 

In terms of the discrete Fourier analysis we are able to analyze the full spectrum of the coupling matrix $\underline{\mathrm H}_\mathrm{SSH}$ for random initial conditions. In Fig.~\ref{fig:FigFourier}(e) we show the result of the frequency spectrum averaged over 10 realizations of random initial conditions as a function of $\delta \lambda$. Remarkably, the spectrum is identical to the eigenspectrum of $ \underline{\mathrm H}_\mathrm{SSH}$, which confirms that the synchronized edge modes are only present in the topological phase ($\delta \lambda>0$). 

\begin{figure}
\begin{center}
\includegraphics[width=0.8\columnwidth]{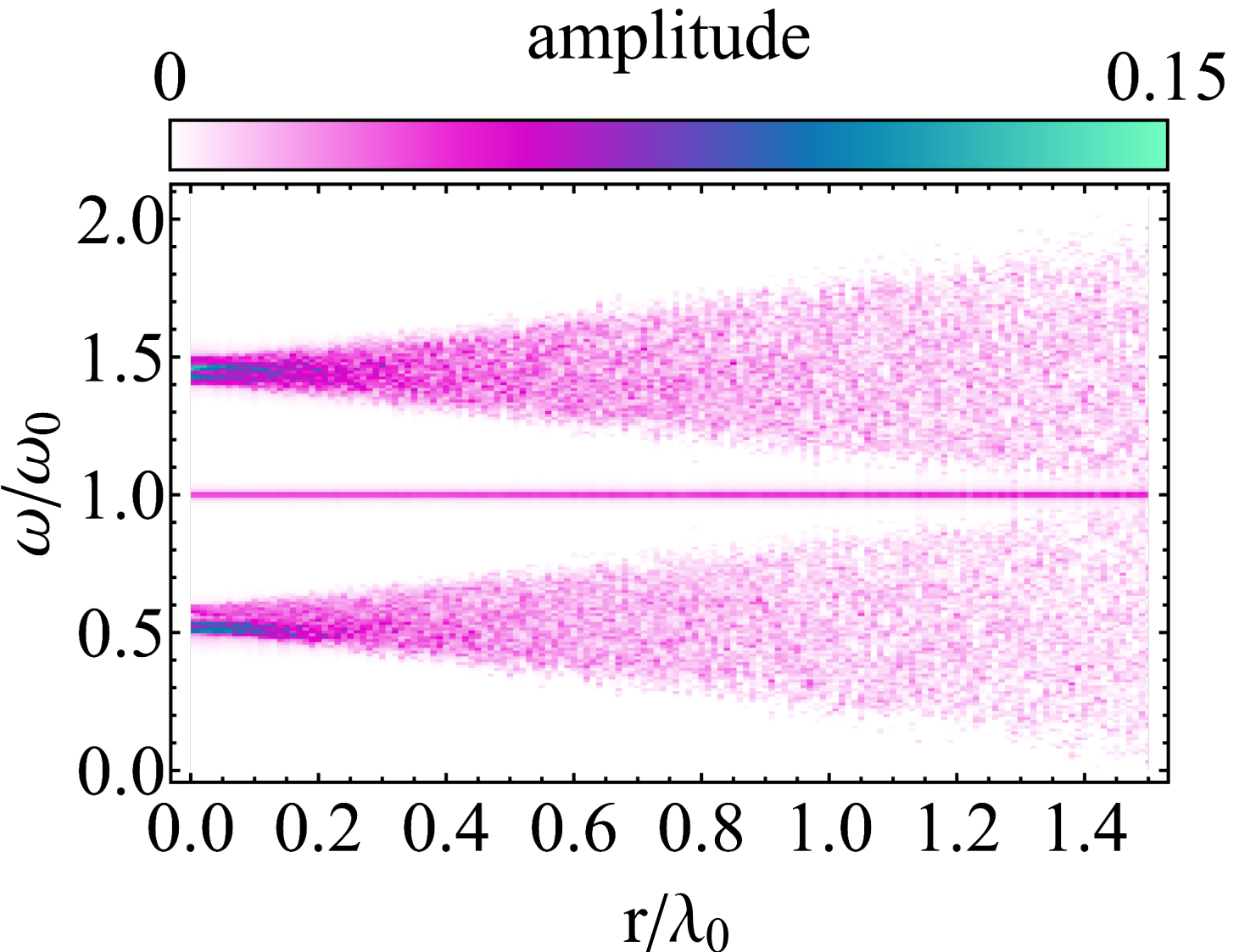}
\end{center}
\caption{Frequency spectrum of the vdP network for $\delta \lambda=0.8\lambda_0$ obtained via discrete Fourier transformation as a function of the disorder strength $r$. While the band frequencies are strongly affected by the disorder, the edge modes located at $\omega = \omega_0$ are robust, even for large amounts of disorder. Parameters: $\kappa_1 = 5\cdot 10^{-3}\omega_0$, $\kappa_2 = 2\kappa_1$, $\lambda _0 = 0.25\omega_0$, $10$ realizations of random initial states for each value of $r$. }
\label{fig:FigDisorder}
\end{figure}

One of the hallmarks of topological insulators is that they exhibit extremely robust surface states since no local perturbation can change their global topology. For the SSH model the robustness arises from an underlying chiral symmetry. To test whether this extraordinary feature is also present in our open nonlinear system of vdP oscillators, we apply random disorder to the couplings between neighboring sites [cf. Eq.~(\ref{eq:SSHHamiltonian})]. In Fig.~\ref{fig:FigDisorder} we show the frequency spectrum of the discrete Fourier analysis for the coupling $\delta \lambda = 0.8t_0$ as a function of the disorder strength $r$ for 10 realizations. While the frequencies within the upper and lower band spread over a wide range as the disorder strength is increased, the edge mode persists even for disorders as large as $r=1.5\lambda_0$ before the bands start overlapping with the zero energy mode.

\subsubsection{Quantum signatures of topological synchronization}
\label{subsec:SSHQuantumFluc}

 \begin{figure*}
\begin{center}
\includegraphics[width=\linewidth]{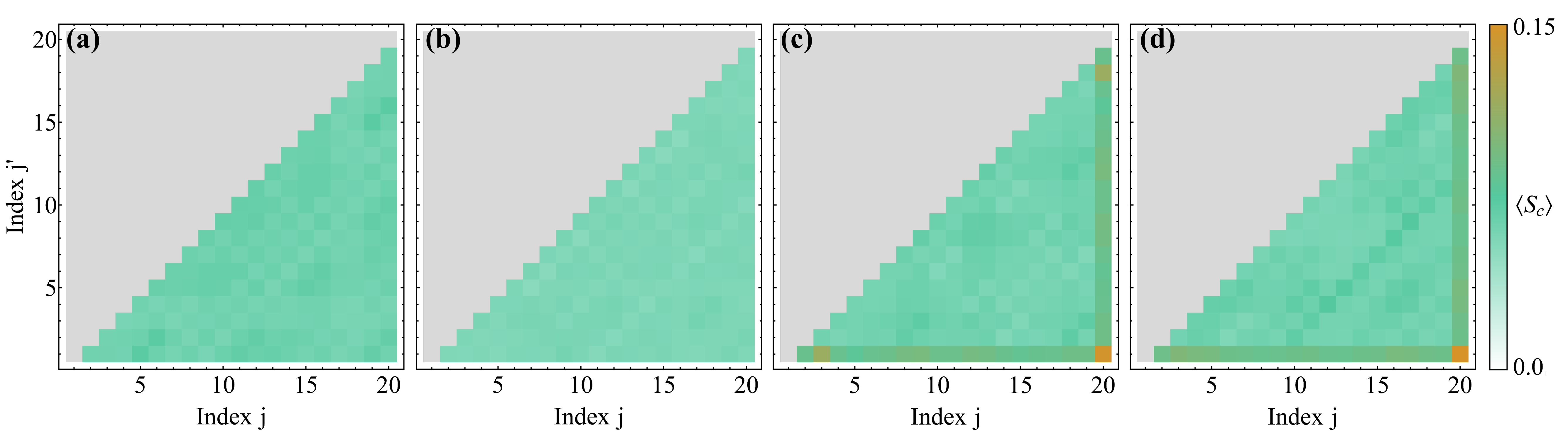}
\end{center}
\caption{
Time-averaged quantum synchronization measure $\left<S_\mathrm c\right>$ between lattice site $j$ and $j'$ of the vdP network corresponding to the mean field dynamics shown in Fig.~\ref{fig:FigFourier}, i.e., with random initial conditions and for different coupling strength (a) $\delta \lambda=-0.8 \lambda_0$, (b) $\delta \lambda = - 0.4\lambda_0$, (c) $\delta \lambda = 0.4 \lambda_0$, and (d) $\delta \lambda = 0.6 \lambda_0$. While in the trivial phase [panels (a) and (b)] there is no  remarkable synchronization between any two lattice sites, in the topological phase the oscillators located at the edges exhibit significantly larger values of synchronization. (e) Quantum synchronization measure $\left<\bar S_\mathrm c(j,j')\right>$ between the two edges of $j=1$ and $j'=20$ in the topological phase ($\delta \lambda =0.8\lambda_0$) as a function of the disorder strength $r$ and averaged over 100 realizations of disorder. The edge state quantum synchronization is topologically protected and robust even for large amounts of disorder. Parameters: $\kappa_1 = 5\cdot 10^{-3}\omega_0$, $\kappa_2 = 2\kappa_1$, $\lambda _0 = 0.25\omega_0$, $\omega_0 t_i = 2\cdot 10^4$, $\omega_0 t_f=2.4\cdot 10^4$.}
\label{fig:FigFlucSSH1Random}
\end{figure*}

In the previous section the focus has been the effects of topology on the mean field dynamics. Thus, the observed signatures are classical in nature even though the underlying model is quantum. By contrast, in this section we analyze the quantum fluctuations about the mean field amplitudes and investigate whether a similar interplay of topology and synchronization carries on beyond the mean field level. To this end, we use the measure $\left<S_\mathrm{c}(j,j')\right>$ [cf. Eq.~(\ref{eq:AvgCompleteSynch})] to quantify quantum synchronization between two lattice sites $j$ and $j'$. For the effective quantum model [cf. Eq.~(\ref{eq:MasterEquationAlpha})--(\ref{eq:EffectiveHamiltonian})] this measure can be calculated via the covariance matrix $\underline{\mathrm C}$ and we use throughout this section the mean field amplitudes after relaxation as initial conditions. We focus here on the case of random initial conditions, that is the quantum fluctuations corresponding to the mean field dynamics shown in Figs.~\ref{fig:FigFourier}(a)--(d). In this way we are able to highlight the topological signatures while keeping the discussion precise. Furthermore, as we show in App.~\ref{secApp:QuantumFlucEV}, the significant results that occur for random initial conditions also carry on if eigenstates $\boldsymbol \alpha^{(l)}$ are chosen as initial conditions (similar as to the topological effects observed in the mean field dynamics). 

In Figs.~\ref{fig:FigFlucSSH1Random}(a)--(d) we show the time-averaged quantum synchronization measure $\left<S_\mathrm c(j,j')\right>$ for different couplings strengths (a) $\delta \lambda=-0.8 \lambda_0$, (b) $\delta \lambda = - 0.4\lambda_0$, (c) $\delta \lambda = 0.4 \lambda_0$, and (d) $\delta \lambda = 0.6 \lambda_0$. A large value of $\left<S_\mathrm c(j,j')\right>$ indicates the existence of quantum synchronization while a small value indicates the lack thereof. In the trivial phase shown in Figs.~\ref{fig:FigFlucSSH1Random}(a) and (b), the synchronization measure is almost uniform with only slight modulations. As there is no  synchronization of the mean field amplitudes for random initial conditions [cf. Figs.~\ref{fig:FigFourier}(a) and (b)], this result is not too surprising. 

A very different scenario is observed for the boundaries of the chain in the topological phase shown in Figs.~\ref{fig:FigFlucSSH1Random}(c) and (d): While for two oscillators in the bulk, $\left<S_\mathrm c(j,j')\right>$ remains similar to the previous trivial phase, the  measure is significantly increased between the two edge oscillators $j'=1$ and $j=20$ (right bottom corner). Hence, the topological synchronization observed for the mean field amplitudes [cf. Figs.~\ref{fig:FigFourier}(c) and (d)] persists even for the effective quantum model and is therefore not a purely classical effect. Note that the quantity is upper bounded by $1$ due to the Heisenberg uncertainty principle [cf. Eq.~(\ref{eq:CompleteSynch})]. Furthermore, significant synchronization can be observed for the oscillators $j'=1$ and $j=3$ as well as $j'=18$ and $j=20$ resulting from the exponential localization of the edge modes at the boundary of the chain. 

Finally, we test the robustness of the edge state synchronization when quantum fluctuations are included. In Fig.~\ref{fig:FigFlucSSH1Random}(e) we show $\left<\bar S_\mathrm c(j,j')\right>$ between the two boundary oscillators ($j=1$ and $j'=20$) in the topological phase with coupling $\delta \lambda =0.8\lambda_0$ as a function of the disorder strength $r$. Here, the overbar denotes that the quantity is averaged over 100 realizations of disorder. Similar as to the results of the mean field dynamics, also the quantum synchronization of the edges is robust for large amounts of disorder. 

\subsection{Example (II): 2D breathing Kagome lattice}
\label{sec:Kagome}

\subsubsection{Edge and corner states in the breathing Kagome lattice}
\label{subsec:KagomeGeneral}

 \begin{figure*}
\begin{center}
\includegraphics[width=\linewidth]{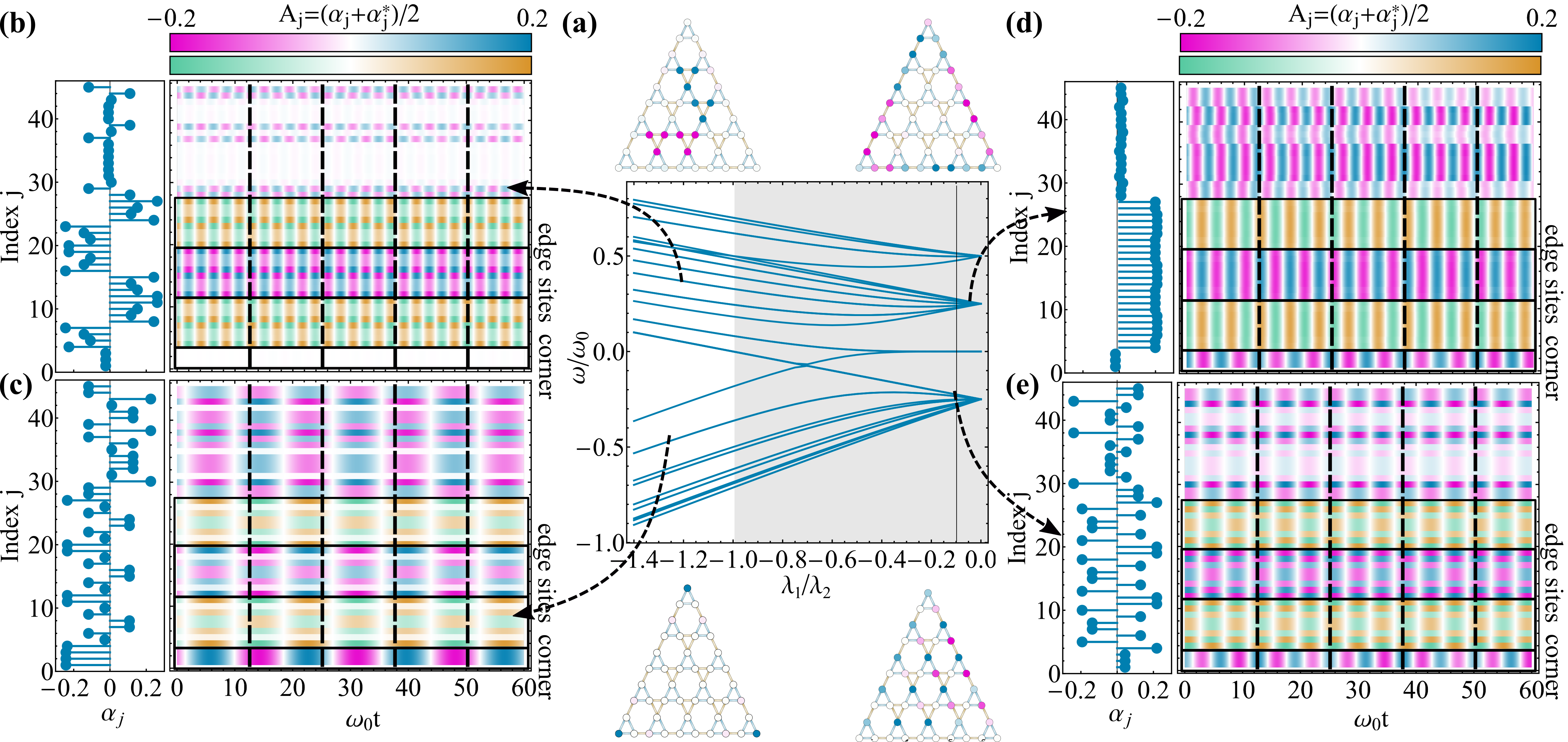}
\end{center}
\caption{(a) Eigenspectrum of the breathing Kagome lattice $\underline{\mathrm H}_\mathrm{Kag}$ as a function of $\lambda_1$ for $5$ upward triangles along each edge, i.e., $N=45$ lattice sites in total. The gray region indicates the nontrivial phase ($\lambda_1/\lambda_2>-1.0$) of the bulk Hamiltonian. The top and bottom lattices show examples of eigenstates for the three bands and the corner states deep in the topological phase $\lambda_1 /\lambda_2 = -0.1$ as indicate by the vertical line. (b)-(e) Left panels show the amplitude $a_j^{(l)}$ of the eigenvectors corresponding to the eigenvalues $\hbar \mu^{(l)}$ marked by the arrows of the eigenspectrum (a), specifically (b) $\mu^{(l)} = 1.7 \lambda_2$ for $\lambda_1 = - 1.2 \lambda_2$, (c) $\mu^{(l)} = -2.5 \lambda_2$ for $\lambda_1 = - 1.3 \lambda_2$, (d) $\mu^{(l)} = 1.0 \lambda_2$ for $\lambda_1 = - 0.05 \lambda_2$ (a strict edge state), (e) $\mu^{(l)} = -1.0 \lambda_2$ for $\lambda_1 = - 0.1 \lambda_2$. These eigenstates are the initial states for the dynamics of $A_j(t)$ of each oscillator $j$ shown in the right panels \new{after the relaxation time $\omega_0 t_\mathrm{rel}$}. The index $j$ of lattices sites is chosen in such a way that $j=1,2,3$ correspond to the the corners of the lattice with dynamics shown in pink/blue color scale ($j=1$ top corner, $j=2$ left corner, $j=3$ right corner), $4\leq j\leq 11$  to the sites along the left edge with dynamics shown in  green/gold color scale ,  $12\leq j\leq 19$  to the sites along the right edge with dynamics shown in pink/blue color scale ,  $20\leq j\leq 27$  to the sites along the bottom edge of the lattice shown green/gold color scale. The remaining indices ($28\leq j\leq 45$) with dynamics shown in pink/blue color scale represent bulk oscillators. Parameters: $\kappa_1 = 5\cdot 10^{-4} \omega_0$, $\kappa_2 = 10^{-2}\omega_0$, $\lambda_2 = 0.25 \omega_0$.}
\label{fig:DynamicsKagome}
\end{figure*}

As many fascinating topological effects occur in dimensions higher than one, we also investigate an example of a two-dimensional system. We consider here the breathing Kagome lattice \cite{ezawa2018higher,kunst2018lattice,bolens2019topological,xue2019acoustic}, which is a natural extension of the SSH chain and a paradigmatic model of a so-called higher-order topological insulator (TI) \cite{schindler2018higher}: While an ordinary $d$-dimensional TI, such as the SSH chain, exhibits $d-1$ dimensional topological edge states, in a higher-order TI $d-n$ dimensional topological boundary states emerge with $n>1$ and the $d-1$ dimensional topological edge states are absent. The reason for this peculiar phenomenon is that the boundary of a higher-order TI itself represents an ordinary TI. For the breathing Kagome lattice with a generalized chiral, time-reversal and particle-hole symmetry (BDI class), the boundary states are zero-dimensional corner states in the topological phase \cite{ni2019observation}. Recently, their robustness has been discussed and it has been realized that they are pinned robustly to zero only if the perturbations respect the generalized chiral and crystalline symmetries and the lattice connectivity \cite{van2020topological,PhysRevB.105.085411}. This is the case considered in this work.

The tight-binding Hamiltonian of the lattice is given by
\begin{equation}
\label{eq:KagomeHamiltonian}
H_\mathrm{Kag}=\hbar\lambda_1 \sum\limits_{\braket{i,j}\in \Delta} a_i^\dagger a_j +\hbar \lambda_2 \sum\limits_{\braket{i,j}\in \nabla} a_i^\dagger a_j, 
\end{equation}
where the two sums are over neighboring sites in the upward and downward triangles, respectively; see Fig.~\ref{fig:DynamicsKagome}(a). Throughout we assume $\lambda_1\leq 0$ and $\lambda_2>0$ as we are only interested in the transition from trivial ($\lambda_1/\lambda_2 <-1$) to topological insulator phase ($\lambda_1/\lambda_2 >-1$). For $\lambda_1/\lambda_2>0$ there exists another phase transition to a metallic phase at $\lambda_1/\lambda_2= 1/2$ \cite{ezawa2018higher}, however, we do not consider this transition in the present work.

In Fig.~\ref{fig:DynamicsKagome}(a) we show the energy spectrum of the breathing Kagome lattice containing $15$ upward triangles as a function of $\lambda_1$, i.e. the coupling within upward triangles. Above and below the spectrum, exemplary eigenstates of the different bands and the band gap are shown for the parameter $\lambda_1/\lambda_2 = -0.1$, i.e. the topological phase: The top band contains only strict bulk states which have (almost) no overlap with the boundary of the lattice as shown in the top left. The bottom band on the other hand contains states delocalized over the lattice  as shown in the bottom right lattice. In contrast, the middle band contains edge states which are localized at the boundary of the lattice as shown in the top right figure. Lastly, the band gap located at zero energy consists of three degenerate states exponentially localized at the corners of the lattice, see the left bottom lattice. The latter are protected by the generalized chiral symmetry. 

\subsubsection{Topological effects on synchronization at the mean field level}
\label{subsec:TopologyKagome}

In the following we investigate the mean field dynamics of the vdP network in the breathing Kagome lattice. To this end, the matrix governing the system dynamics without dissipation  in Eq.~(\ref{eq:VdPLattice}) is given by
\begin{equation}
\label{eq:ClosedSystemKagome}
\underline{\mathrm H} = \hbar\omega_0 \underline{1} + \underline{\mathrm H}_\mathrm{Kag},
\end{equation}
where $\underline{\mathrm H}_\mathrm{Kag}$ denotes the matrix corresponding to the Hamiltonian defined in Eq.~(\ref{eq:KagomeHamiltonian}). 

Following our general analytic considerations and the results of the previous Example (I), we expect that the oscillatory dynamics is dominated by the closed system dynamics given by Hamiltonian~(\ref{eq:ClosedSystemKagome}). In Figs.~\ref{fig:DynamicsKagome}(b)--(e) we show the amplitude $A_j(t)$ of each oscillator $j$ as a function of time with eigenstates $\boldsymbol \alpha^{(l)}$ as initial states; shown in the left panels of Figs.~\ref{fig:DynamicsKagome}(b)--(e). We start our discussion in the trivial phase shown in Figs.~\ref{fig:DynamicsKagome}(b) and (c), where the total vdP network is synchronized, oscillating with frequency $\omega_0 +\mu^{(l)}$  [a larger frequency in panel (b) where $\mu^{(l)} > 0$ and a smaller frequency in panel (c) where $\mu^{(l)}< 0$] and oscillation amplitudes proportional to the chosen initial eigenstate $\boldsymbol \alpha^{(l)}$. As in the previous Example (I) 
the dynamics in the trivial phase is thus  given by $\boldsymbol \alpha(t) = c \boldsymbol \alpha^{(l)} \exp\{-i[\omega+\mu^{(l)}]t\}$.

In Figs.~\ref{fig:DynamicsKagome}(d) and (e) we show examples of the dynamics in the topological phase. If the initial eigenstate is completely delocalized across the lattice [cf. Fig.~\ref{fig:DynamicsKagome}(e)], the oscillator dynamics follows the same principles as discussed previously, i.e. complete synchronized oscillations with common frequency $\omega_0 + \mu^{(l)}$ except of the three corners (index $j=1,2,3$), which amplitudes oscillate with the intrinsic frequency $\omega_0$. The latter additionally exhibit a temporal amplitude modulation as a result of the finite amplitude of the initial state at the corners, i.e. $A_j(t) = \bar A_j \cos\{[\omega_0 + \mu^{(l)}]t\} + \bar A_\mathrm{corner}\cos(\omega_0 t)$, where $\bar A_j$ and $\bar A_\mathrm{corner}$ reflect the mixture of the two participating frequencies. 

However, if the initial eigenstate is localized at the edges as shown in Fig.~\ref{fig:DynamicsKagome}(d) we additionally observed excitation of all bulk lattice oscillating with a common frequency. However, their frequency does not match the corner nor the edge oscillators. Moreover, their amplitude is not constant but exhibits modulations over time. The reason for this effect is that the top band in Fig.~\ref{fig:DynamicsKagome}(a) consists of strict bulk states without participation of any edge or corner lattice sites. Due to the nonlinearity in the dynamics, these strict bulk states may grow, yet, as a superposition of many of those leading to the observed amplitude modulations. 

 \begin{figure*}
\begin{center}
\includegraphics[width=\linewidth]{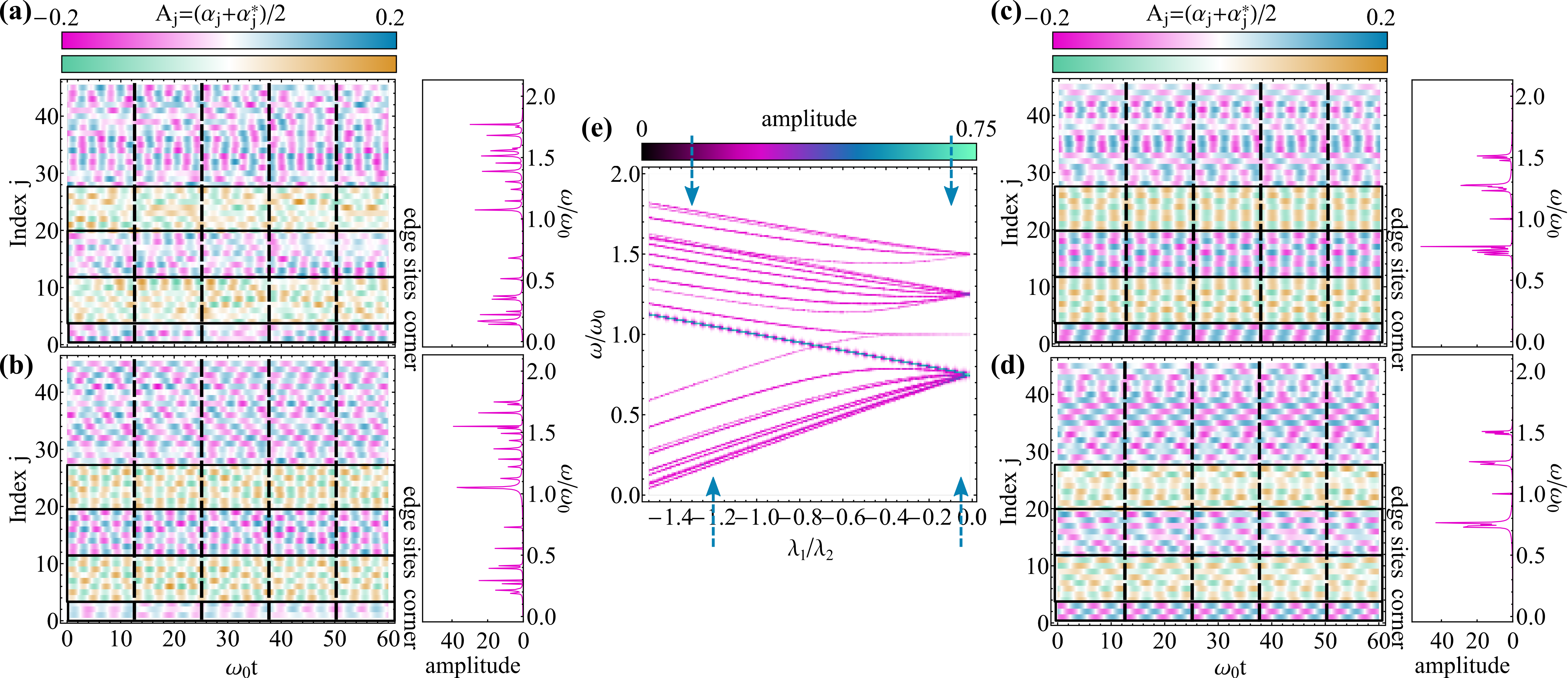}
\end{center}
\caption{(a)--(d) Left panels show the amplitude dynamics $A_j(t)$ of oscillator $j$ -- where we choose an ordering according to Fig.~\ref{fig:DynamicsKagome} and two color scales to differentiate corner, different edge and bulk lattice sites -- in the vdP Kagome network \new{after the relaxation time $\omega_0 t_\mathrm{rel}$} with random initial conditions for different couplings: (a) $\lambda_1/\lambda_2 = -1.3$, (b) $\lambda_1/\lambda_2 = -1.2$, (c) $\lambda_1/\lambda_2 = -0.1$, and (d) $\lambda_1 / \lambda_2 = -0.05$ [also marked by the blue arrows in panel (e)]. The  right panels show the frequency spectrum of the dynamics shown in the left panels obtained via discrete Fourier analysis. (e) Frequency spectrum \new{$H_0 + H_\mathrm{Kag}$ (or equivalently $\underline{\mathrm{H}}_0 + \underline{\mathrm{H}}_\mathrm{Kag}$)} obtained from the oscillator dynamics for random initial conditions as function of the upward-triangle coupling $\lambda_1$, where for each coupling 10 realizations of initial conditions are averaged. Parameters: $\kappa_1 = 5\cdot 10^{-4} \omega_0$, $\kappa_2 = 10^{-2}\omega_0$, $\lambda_2 = 0.25 \omega_0$.}
\label{fig:DynamicsKagomeFourier1}
\end{figure*}

We are now interested whether the topological synchronization with random initial conditions also carries over to the two dimensional model. In Figs.~\ref{fig:DynamicsKagomeFourier1}(a)--(d) we show in the left panels the oscillator dynamics $A_j(t)$ as function of time for different coupling strengths with ordering of the oscillators according to Fig.~\ref{fig:DynamicsKagome}. In the corresponding right panels we show the frequency spectrum of the particular realization of oscillator dynamics obtained via discrete Fourier transformation. In the trivial phase shown in Figs.~\ref{fig:DynamicsKagomeFourier1}(a) and (b) clear signatures of synchronization are again missing and the dynamics is governed by a randomly distributed superposition of many eigenstates oscillating with different frequencies. The emergent complex oscillation pattern may exhibit temporal synchronized structures, however, these vanish again over longer times.  

By contrast in the topological phase, shown in Figs.~\ref{fig:DynamicsKagomeFourier1}(c) and (d), the dynamics of the network appears more structured and synchronized. The reason for this is that fewer frequencies are available in the eigenspectrum of $\underline{\mathrm H}_\mathrm{Kag}$, and especially in the specific realization that there exists a dominant frequency as indicated by the large peak in the frequency spectra (see right panels). However, most notably the oscillators located at the three corners of the lattice ($j=1,2,3$) are phase locked [cf. Eq.~(\ref{eq:SynchPhiCondition})] and oscillate with the intrinsic frequency $\omega_0$. Thus, as in the SSH chain, the topological character of the lattice is reflected in the amplitude oscillations even for random initial conditions. This also allows us to reconstruct the full eigenspectrum of the topological coupling matrix $\underline{\mathrm H}_\mathrm{Kag}$ via discrete Fourier analysis of oscillator dynamics, which we show in Fig.~\ref{fig:DynamicsKagomeFourier1}(e) averaged over 10 realizations of initial conditions. \new{The original eigenspectrum of $\underline{\mathrm H}_\mathrm{Kag}$ has a highly degenerate eigenvalue spanning diagonally across, which is also observed in the reconstruction of the frequency spectrum as much larger amplitude (blue) than the other eigenvalues (pink).}

\begin{figure}
\begin{center}
\includegraphics[width=0.8\columnwidth]{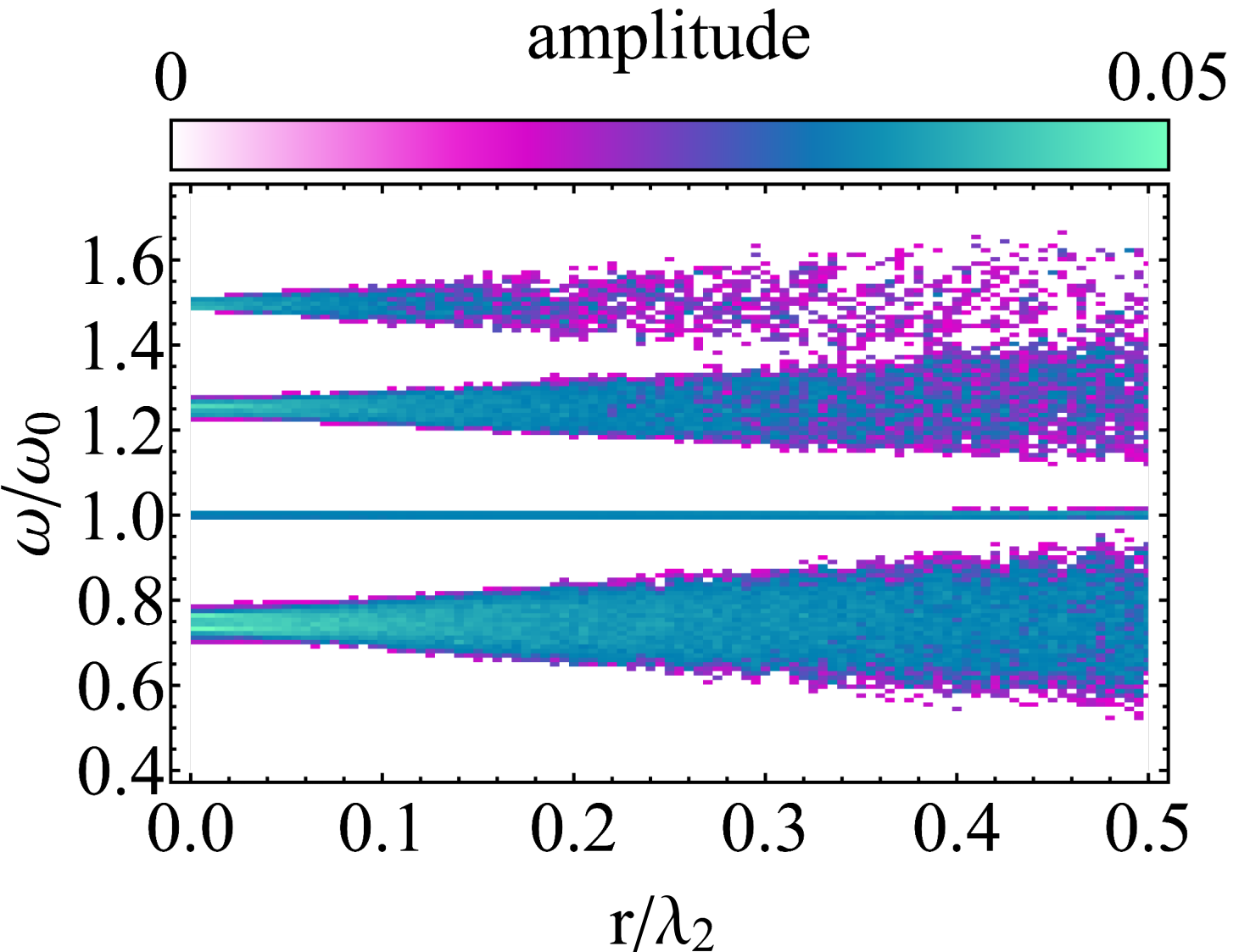}
\end{center}
\caption{Frequency spectrum of the vdP network for $\lambda_1=-0.05\lambda_2$ obtained via discrete Fourier transformation as a function of the disorder strength $r$. While the band frequencies are strongly affected by the disorder, the corner modes located at $\omega = \omega_0$ are robust, even for large amounts of disorder. Parameters: $\kappa_1 = 5\cdot 10^{-4} \omega_0$, $\kappa_2 = 10^{-2}\omega_0$, $\lambda_2 = 0.25 \omega_0$, $10$ realizations of random initial states for each data value of $r$.}
\label{fig:FigDisorderKagome1}
\end{figure}

Lastly, we test the robustness of the observed corner state synchronization. As the corner states of the breathing Kagome lattice are protected by a generalized chiral symmetry, we expect protection of the corner state synchronization. In Fig.~\ref{fig:FigDisorderKagome1} we show the frequency spectrum of the discrete Fourier analysis for the coupling $\lambda_1 = -0.05 \lambda_2$ as a function of the disorder strength $r$ for 10 realizations. The zero-energy corner modes are robust against disorder strengths as large as $r=0.4 \lambda_2$ before they are affected by the perturbations. In contrary, the bands spread over a wider range as the disorder strength is increased. Furthermore, this analysis also shows that the edge states are \textit{not} topologically protected in the higher order TI. 

\subsubsection{Quantum signatures of topological synchronization}
\label{subsec:KagomeQuantumFluc}

 \begin{figure*}
\begin{center}
\includegraphics[width=\linewidth]{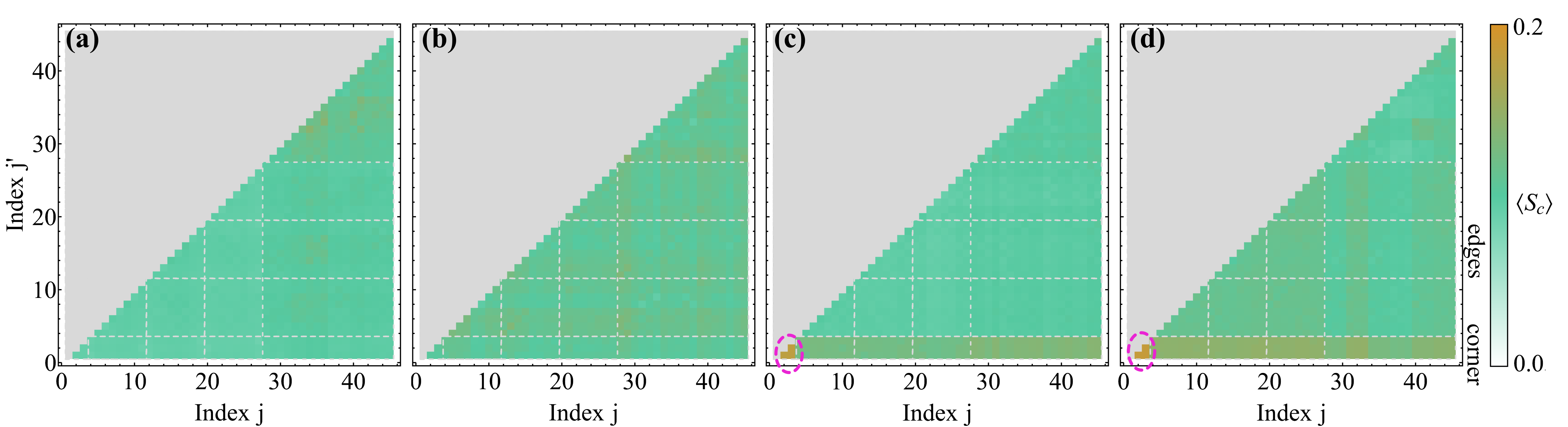}
\end{center}
\caption{Time-averaged quantum complete synchronization measure $\left<S_\mathrm c\right>$ between lattice site $j$ and $j'$ of the vdP Kagome network corresponding to the mean field dynamics shown in Fig.~\ref{fig:DynamicsKagomeFourier1}, i.e., with random initial conditions and for different coupling strength (a) $\lambda_1/\lambda_2 = -1.3$, (b) $\lambda_1/\lambda_2 = -1.2$, (c) $\lambda_1/\lambda_2 = -0.1$, and (d) $\lambda_1 / \lambda_2 = -0.05$. The ordering of the lattice sites is the same as in Fig.~\ref{fig:DynamicsKagomeFourier1} especially $j=1,2,3$ correspond to the corners of the lattice. While in the trivial phase [panels (a) and (b)] there is no clear pattern of synchronization between any two lattice sites, in the topological phase the oscillators located at the corners (encircled in pink) exhibit significantly larger values of synchronization. (e) Quantum synchronization measure $\left<\bar S_\mathrm c(j,j')\right>$ between the three corners [($j=1,j'=2$), ($j=1,j'=3$) and ($j=2,j'=3$)] in the topological phase ($\lambda_1/\lambda_2 =-0.05$) as a function of the disorder strength $r$ and averaged over 100 realizations of disorder. The corner state quantum synchronization is protected and robust even for large amounts of disorder. Parameters: $\kappa_1 = 5\cdot 10^{-4} \omega_0$, $\kappa_2 = 10^{-2}\omega_0$, $\lambda_2 = 0.25 \omega_0$, $\omega_0 t_i = 2 \cdot 10^4$, $\omega_0 t_f=2.3\cdot 10^4$.}
\label{fig:FigFlucKagomeRandomSC}
\end{figure*}

After we have observed in the previous section that topological protected synchronization of the mean field amplitudes also exists for a higher-order TI, we now analyze its quantum fluctuations  quantified in terms of $\left<S_\mathrm{c}(j,j')\right>$ [cf. Eq.~(\ref{eq:AvgCompleteSynch})] between two lattices sites $j$ and $j'$. Similar as to the case of the SSH model we choose the mean field amplitudes after relaxation as initial conditions and focus on the case of random initial conditions corresponding to the dynamics shown in Figs.~\ref{fig:DynamicsKagomeFourier1}(a)--(d). 

In Figs.~\ref{fig:FigFlucKagomeRandomSC}(a)--(d) we show the time-averaged quantum synchronization measure $\left<S_\mathrm c(j,j')\right>$ for different couplings strengths (a) $\lambda_1/\lambda_2 = -1.3$, (b) $\lambda_1/\lambda_2 = -1.2$, (c) $\lambda_1/\lambda_2 = -0.1$, and (d) $\lambda_1 / \lambda_2 = -0.05$. The ordering of the lattice sites corresponds to Fig.~\ref{fig:DynamicsKagomeFourier1}, especially $j',j=1,2,3$ are the corners of the lattice. We are mostly interested in the latter as they are expected to show quantum signatures of synchronization. In accordance with the classical mean field amplitudes of Figs.~\ref{fig:DynamicsKagomeFourier1}(a) and (b), the synchronization measure is almost uniform in the trivial phase shown in Figs.~\ref{fig:FigFlucKagomeRandomSC}(a) and (b). 

However, in the topological phase shown in Figs.~\ref{fig:FigFlucKagomeRandomSC}(c) and (d) we observe that the oscillators located the corners $(j',j=1,2,3)$ are significantly synchronized with one another as indicated by the large value of $\left<S_\mathrm c(j,j')\right>$ at the left bottom corner (highlighted by the pink dashed circle). As a reminder the quantity is bounded by $1$ [cf. Eq.~(\ref{eq:CompleteSynch})]. Moreover, the quantum synchronization measure indicates that all three corner oscillators are synchronized with the edges and parts of the bulk. This feature is also present for the two edges of the SSH model in Example (I) [cf. Fig.~\ref{fig:FigFlucSSH1Random}]. By contrast, for any two bulk oscillators, $\left<S_\mathrm c(j,j')\right>$ remains similar to the previous trivial phase. 

Finally, we test the topological protection of the corner state synchronization at the quantum level. In Fig.~\ref{fig:FigFlucKagomeRandomSC}(e) we show $\left<\bar S_\mathrm c(j,j')\right>$ between the corners [($j=1,j'=2$), ($j=1,j'=3$) and ($j=2,j'=3$)] in the topological phase ($\delta \lambda =0.8\lambda_0$) as a function of the disorder strength $r$. Here, the overbar denotes the average over 100 realizations of disorder. Consistent with the previous observations, also the quantum synchronization of the corners is robust for large amounts of disorder. 

\section{Discussion}
\label{sec:Discussion}

An adequate formulation of topological insulators can be provided within the framework of solid state band theory and it is thus far from obvious whether and how their effects persist if affected by dissipation. Moreover, in the case investigated in this work dissipation  represents a necessary resource to drive the system far away from equilibrium. Remarkably, despite the tremendous consequences open system conditions in combination with nonlinearities can have on the system dynamics, our examples show that topological features remain in such a scenario, which allows us to utilize them in our favor. 

\new{The observed topological boundary synchronization at the mean field level even if an eigenstate of the topological Hamiltonian is chosen as initial state is rather surprising and represents a key difference to systems without dissipation: In closed system dynamics, an  initial eigenstate of the Hamiltonian will persist and will not mix with other eigenstates as they are orthogonal to each other. However, we observe in the nonequilibrium system that in the topological phase the edge oscillators are excited and their amplitudes grow, even though the initial eigenstate of the system Hamiltonian has very little amplitude at the edges. This may be understood as follows: Deep in the topological phase the oscillators located at the edges/corners are only weakly coupled to the rest of lattice. As their initial amplitude  is finite it serves as a small perturbation away from their unstable fixed points, and since there exist an eigenstate localized at the edges this eigenstate will eventually grow. By contrast, oscillators in the bulk with small initial amplitudes are prevented from being excited because they require many other bulk oscillators to change their amplitudes accordingly. Interestingly, if the initial eigenstate is highly localized at the boundaries, they also serve as a perturbation for the bulk oscillators such that a combination of many bulk modes may be excited; see Fig.~\ref{fig:DynamicsKagome}(d). We therefore argue, that the observed topological boundary synchronization at the mean field level is not restricted to the two specific examples investigated in this work, but also applies to other topological insulator lattices as the topologically protected edge modes are highly localized and all other bulk modes only have a small amplitude at the boundaries.}
Moreover, \new{the synchronized edge modes} inherit the topological protection known from closed systems with remarkable robust dynamics against local (symmetry-preserving) disorder and even random initial conditions. \new{However, disorder in the natural frequencies of the oscillators on the considered lattices will destroy the observed edge synchronization. This is similar to topological insulators under closed system conditions, where the edge states are not robust against perturbations which break the underlying symmetries.}

Often when fluctuations are considered synchronization is lost. Our results, however, show signatures of boundary synchronization beyond the classical mean field approximation. Furthermore, it remains unaffected for large amounts of local disorder in the couplings due to the underlying topology, which has the advantage that even if perfect fabrication of the lattice is impossible our findings can still be observed. This makes our results appealing for experimental realizations, two of which we discuss in more detail in the next section. 

Lastly, let us highlight the ability to reconstruct the full eigenspectrum of the  underlying topological lattice from the oscillator (mean field) dynamics alone. In our numerical studies with only 10 realizations the full spectra of the (closed) SSH and Kagome lattice could already be obtained. This holds the potential of a new experimental mechanism to measure eigenspectra of a topological systems from the  dynamics of a nonlinear system far away from equilibrium and without preparation of an initial state.

\section{Experimental proposal}
\label{sec:Expoeriments}

There are many possible experimental platforms where topological synchronization may be observed in the mean field (classical) as well as the quantum regime. To realize the dynamics of a single vdP oscillator in a quantum system two different implementations have been suggested: The first one uses trapped ions \cite{Lee2013,PhysRevLett.112.094102} as an experimental platform while the second one focuses on optomechanics \cite{PhysRevLett.112.094102,Walter2015}. Both of these proposals bear the potential of vdP oscillator networks and thus to observe the previously discussed topological synchronization phenomena. Therefore, we summarize both of them and discuss their extensions towards topological lattices. 

\begin{figure}
\begin{center}
\includegraphics[width=\columnwidth]{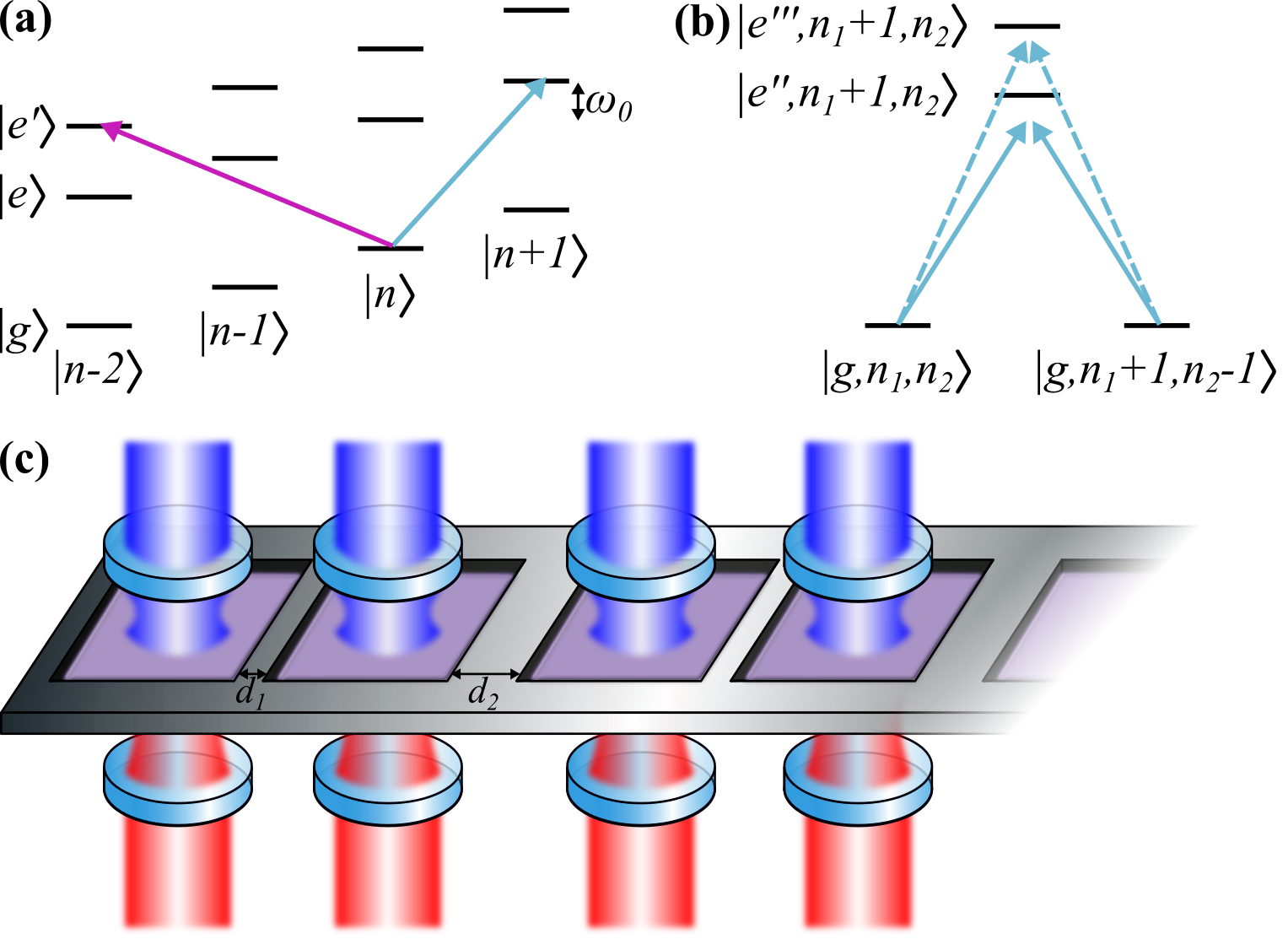}
\end{center}
\caption{Potential experimental realizations of topologically coupled vdP oscillators. (a) and (b) Trapped ions implementation: (a) Level scheme for an ion with trap frequency $\omega_0$. By exciting the blue sideband of the transition $\ket{g}\to\ket{e}$ negative damping may be realized (blue arrow). Simultaneously driving the double red sideband transition $\ket{g}\to \ket{e'}$ allows one to realize the nonlinear damping (pink arrow). (b) Two modes may be coupled via off-resonant excitation of blue sidebands of $\ket{g}\to\ket{e''}$ or $\ket{g}\to\ket{e'''}$, which realizes the alternating couplings between nearest neighbors. (c) Optomechanical implementation: The so-called `membrane in the middle'-setup allows to realize a vdP oscillator with nonlinear damping via a laser detuned to the red two-phonon sideband and negative damping by a laser on the blue one-phonon sideband. Mechanical coupling with alternating distances then allows to implement the topological lattice interactions.}
\label{fig:Exp1}
\end{figure}

A motional mode of a trapped ion in the Lamb-Dicke regime and when the trapping potentials are tight with ground state $\ket{g}$ and excited state $\ket{e}$ is represented as an harmonic oscillator with frequency $\omega_0$. In order to fulfill the required conditions, i.e., negative and nonlinear damping, two side bands are excited simultaneously \cite{leibfried2003quantum}: The first laser drives the transition $\ket{g}\to \ket{e}$ but blue detuned by $\omega_0$ which absorbs one phonon after subsequent decay to the ground state [see Fig.~\ref{fig:Exp1}(a)]. The second laser is double red detuned by $-2\omega_0$ and excites to a state $\ket{e'}$ such that after relaxation to the ground state two phonons have been emitted. The combination of these two processes approximately implement the dissipators of each oscillator in Eq.~(\ref{eq:QauntumModel}). Specific parameters for an implementation with ${^{171}\mathrm{Yb}}^+$ are provided in Ref.~\cite{Lee2013} and references therein.

Additionally, the bosonic modes of neighboring trapped ions need to be coupled: By off-resonantly exciting an additional level $\ket{e''}$ via a blue sideband [see Fig.~\ref{fig:Exp1}(b)] an effective Hamiltonian $H=\hbar\lambda_1 (a_j^\dagger a_{j+1} + a_{j+1}^\dagger a_j)$ for the phonon exchange between oscillator $j$ and $j+1$ may be implemented. Using the same strategy but with an additional level $\ket{e'''}$ leads to a different coupling strength $\lambda_2$. In this way the alternating interactions necessary for topological effects to emerge may be realized in trapped ion experiments. 

Another experimental setup where our proposal could be implemented is provided by optomechanics as sketched in Fig.~\ref{fig:Exp1}(c): The so-called `membrane-in-the-middle'-setup \cite{thompson2008strong,sankey2010strong,aspelmeyer2014cavity} allows to realize two-phonon processes \cite{nunnenkamp2010cooling} as the cavity mode is parametrically modulated by the \textit{squared} position of a movable membrane. The Hamiltonian of a single mechanical membrane inside a cavity driven on the red two-phonon resonance (in the good-cavity limit and focusing on the resonant terms) is given by $H_\mathrm{opt}=\hbar g(a^\dagger a^\dagger b + \mathrm{h.c.})$ \cite{nunnenkamp2010cooling}. The dynamics of the coupled system is described by the master equation
\begin{equation}
\label{eq:OptomechanicsME}
\dot \varrho = -\frac{\mathrm i}{\hbar} \left[H_\mathrm{opt},\varrho\right] + \gamma_\mathrm{c}\mathcal D\left[b\right]\varrho + \gamma(1+\bar n)\mathcal D\left[a\right]\varrho + \gamma \bar n \mathcal D\left[a^\dagger\right] \varrho, 
\end{equation}
where $\bar n$ is the thermal phonon number. The Heisenberg equation of motion for the cavity  is $\dot{b}=-\mathrm{i}gaa-\gamma_\mathrm{c}b/2$. Adiabatically eliminating the optical mode and defining $\kappa_2 \equiv g^4/\gamma_\mathrm c$ results in the two-phonon dissipator of Eq.~(\ref{eq:QauntumModel}). However, the other two linear dissipators in Eq.~(\ref{eq:OptomechanicsME}) are related via the thermal phonon occupation $\bar n$, such that one-phonon gain cannot be larger than the one-phonon loss. Therefore, another laser driving the cavity on the blue one-phonon sideband is needed \cite{PhysRevLett.112.094102, Walter2015}. Nevertheless, compared to Eq.~(\ref{eq:QauntumModel}) an additional one-phonon loss term $\gamma (1+\bar n) \mathcal D[a]\varrho$ is present. If we assume $\gamma (1+\bar n) \ll \kappa_1$, the influence of this additional term can be completely neglected at the mean field level as it only results in an effective damping rate $\tilde \kappa_1 = \kappa_1 - \gamma (1+\bar n)$. Also our results in the quantum regime remain unchanged if one would include such a linear loss channel (see App.~\ref{secApp:LinearLoss}). 

When two membranes share a common support they are mechanically coupled expressed via the Hamiltonian $H = \tilde \lambda_j x_j x_{j+1}$ where $x_j = \sqrt{\hbar/2 m \omega_0}(a_j^\dagger + a_j)$ with oscillator mass $m$. After the rotating-wave approximation and with the definition $\lambda_j = 2 m \omega_0\tilde \lambda_j $, the Hamiltonian becomes equivalent to the coupling Hamiltonian (\ref{eq:GeneralHamiltonian}). The interaction strengths between two oscillators may be modified by altering the distance between neighboring membranes; see Fig.~\ref{fig:Exp1}(c) for an example of the SSH model with staggered distances $d_1$ and $d_2$. Thus, also optomechanics has the potential to observe topologically protected synchronization in a network of quantum vdP oscillators. 

\section{Conclusions}
\label{sec:Conclusions}
We investigate the interplay of topology and synchronization in a network of coupled quantum van der Pol oscillators simulating different topological insulator lattices. We show via a linear stability analysis that the dynamics of the resulting topological lattice of oscillators at the mean field level is governed by the eigenvalues of the topological Hamiltonian and thus reflects the features of the underlying topology even though the system is highly nonlinear and far away from equilibrium. Furthermore, we derive an effective quantum model which takes quadratic quantum fluctuations about the classical trajectories into account in order to investigate quantum signatures of topological synchronization beyond the mean field approximation. For two specific examples of topological insulator models in one and two dimensions we demonstrate that in the nontrivial phase, synchronization at the boundaries is always present independent of initial conditions, and that it inherits the protection against perturbations from the underlying lattice structure. In terms of a Fourier analysis of the oscillations, we are able to reconstruct the full topological eigenspectrum of the system, which not only represents a possible route to observe topological synchronization of boundary modes experimentally, e.g. in trapped ions or optomechanics, but an additional opportunity to measure topological eigenspectra solely from the dynamics of a highly nonlinear and open system. 

Researchers and engineers make great efforts to fabricate dynamical systems which are nearly identical in order to facilitate the emergence of synchronized collective behavior in large networks. However, our work demonstrates a general advantage of \new{topological lattices} in the design of potential experiments and devices as fabrication errors and longterm degradation are circumvented in this way. This is especially important in networks where specific nodes need special protection. While the examples investigated in this work posses zero dimensional protected boundary states, our work can easily be extended to host higher dimensional topologically protected states for  additional robust network nodes. Synchronization is desirable in situations where high oscillating power, strong coherence, or low phase noise are needed, such as lasers \cite{thornburg1997chaos}, phase-locked loops \cite{lynch1995mode}, Josephson junction arrays \cite{cawthorne1999synchronized,fazio2001quantum}, spin-torque resonators \cite{slavin2009spin}, quantum heat engines \cite{jaseem2020quantum} or power grids \cite{Nishikawa_2015}. Even today, the originally observed phenomenon of clock synchronization remains a crucial application for modern communication networks \cite{bellamy1995digital,narula2018requirements} and has recently been extended to quantum networks and quantum key distribution protocols \cite{calderaro2020fast,agnesi2020simple}. All of these examples require the synchronized behavior to be robust  to fulfill their desired purpose and will benefit from the application of topology. Given the universality of the concept of combining nonlinear dynamics in open quantum systems with topological phases of matter, we expect that our approach could be successfully applied also to other systems where robust dynamics is crucial.

\begin{acknowledgments}
We thank M. T. Eiles and W. Munro for invigorating discussions and for insightful comments on the manuscript. C.W.W. acknowledges support from the Max-Planck Gesellschaft via the MPI-PKS Next Step fellowship and is financially supported by the Deutsche Forschungsgemeinschaft (DFG, German Research Foundation) – Project No. 496502542 (WA 5170/1-1). G.P. acknowledges support
from the Spanish Ministry of Science and Innovation through the grant No. PID2020-117787GB-100 and from the CSIC Research Platform on Quantum Technologies PTI-001.
\end{acknowledgments}

\appendix

\section{Derivation of the effective quantum model in the comoving frame}
\label{secApp:EffectiveQM}

In this section we derive the effective master Eq.~(\ref{eq:MasterEquationAlpha}) of the main manuscript governing the dynamics of quantum fluctuations\new{, which follows the derivations found in Ref.~\cite{PhysRevX.4.011015,bastidas2015}.} Our starting point is the master equation (throughout this section we set $\hbar\equiv 1$)
\begin{equation}
\dot{\varrho} = -\mathrm i\left[H_\mathrm S,\varrho\right] +  \sum\limits_{j} \left\{\kappa_1\mathcal D\left[a_j^\dagger\right]\varrho +\kappa_2\mathcal D\left[a_j^2\right]\varrho \right\}. 
\end{equation}
with $H_\mathrm S = \omega_0 \sum_j a_j^\dagger a_j +H_\mathrm{top}$. The density matrix in the displaced frame is defined in terms of the displacement operator $\varrho_\alpha(t)= \Ddalpha \varrho(t) \Dalpha$ with corresponding master equation:
\begin{equation}
\label{eq:MasterEquationAlphaSM}
\begin{aligned}
\dot \varrho_\alpha &= -\mathrm i \left[\tilde H_\mathrm S,\varrho_\alpha\right] + \mathcal L_1\varrho_\alpha + \mathcal L_2 \varrho_\alpha 
\end{aligned}
\end{equation}
where the transformed system Hamiltonian is given by
\begin{equation}
\begin{aligned}
\tilde H_\mathrm S &=\Ddalpha\left(H_\mathrm S - \mathrm i \partial_t \right) \Dalpha,\\
\end{aligned}
\end{equation}
and the two dissipators by 
\begin{equation}
\begin{aligned}
\mathcal L_1\varrho_\alpha&=\kappa_1 \sum\limits_j \Ddalpha\mathcal D\left[a_j^\dagger\right]\varrho\Dalpha,\\
\mathcal L_2 \varrho_\alpha &=   \kappa_2 \sum\limits_j \Ddalpha\mathcal D\left[a_j^2\right]\varrho\Dalpha.
\end{aligned}
\end{equation}

We now evaluate the effect of applying the displacement operator on the different terms appearing in Eq.~(\ref{eq:MasterEquationAlphaSM}). For transformed Hamiltonian $\tilde H_\mathrm S$ we obtain

\begin{widetext}
\begin{equation}
\begin{aligned}
\tilde H_\mathrm S =& \omega_0 \sum\limits_j \Ddalpha a_j^\dagger a_j \Dalpha + \Ddalpha H_\mathrm{top} \Dalpha -\mathrm i \Ddalpha \partial_t \Dalpha\\
=& H_\mathrm S + H_\mathrm{top} + \omega_0\sum\limits_j \left( \alpha_j a_j^\dagger + \alpha_j^\ast a_j + \left|a_j\right|^2\right)  + \sum\limits_{jj'} \left\{\lambda_{jj'}\left(\alpha_j^\ast a_{j'} + \alpha_{j'} a_j^\dagger + \alpha_j^\ast a_{j'}\right) + \lambda^\ast_{j'j} \left(\alpha_{j'}^\ast a_j + \alpha_j a_{j'}^\dagger+ \alpha_{j'}^\ast a_j\right)\right\} \\
&- \mathrm i \sum\limits_j \left[\dot \alpha_j \left( a_j^\dagger + \frac{1}{2}\alpha_j^\ast\right) -\dot \alpha_j^\ast \left(a_j+ \frac{1}{2}\alpha_j \right)\right].
\end{aligned}
\end{equation}
For the dissipative term proportional to $\kappa_1$ we obtain
\begin{equation}
\mathcal L_1\varrho_\alpha  = \kappa_1 \sum\limits_j \Ddalpha \mathcal D\left[a_j^\dagger\right]\varrho \Dalpha  = \kappa_1 \sum\limits_j \left\{\mathcal D\left[\alpha_j^\dagger\right] \varrho_\alpha -\mathrm i \frac{1}{2}\left[\mathrm i \alpha_j a_j^\dagger - \mathrm i \alpha_j^\ast a_j,\varrho_\alpha \right]\right\},
\end{equation}
and for the dissipative term proportional to $\kappa_2$ we obtain
\begin{equation}
\begin{aligned}
\mathcal L_1\varrho_\alpha  =& \kappa_2 \sum\limits_j \Ddalpha \mathcal D\left[a_j^2\right]\varrho \Dalpha \\
=& \kappa_2 \sum\limits_j \left\{\mathcal D\left[ a_j^2\right]\varrho_\alpha +4\left|\alpha_j\right|^2\mathcal D\left[ a_j\right]\varrho_\alpha + 2\alpha_j^\ast \left(a_j^2\varrho_\alpha a_j^\dagger -\frac{1}{2}a_j^2 a_j^\dagger \varrho_\alpha  -\frac{1}{2} \varrho_\alpha a_j^2 a_j^\dagger\right)  \right. \\
&\left. + 2\alpha_j \left(a_j\varrho_\alpha (a_j^\dagger)^2 -\frac{1}{2}a_j (a_j^\dagger)^2 \varrho_\alpha  -\frac{1}{2} \varrho_\alpha a_j (a_j^\dagger)^2\right)+ \frac{1}{2}\left[(\alpha_j^\ast)^2 a_j^2 - \alpha_j^2 (\alpha_j^\dagger)^2,\varrho_\alpha\right] + \left|\alpha_j\right|^2\left[\alpha^\ast_j a_j - \alpha_j a_j^\dagger, \varrho_\alpha\right] \right\}\\
=& \kappa_2 \sum\limits_j \left\{ 4\left|\alpha_j\right|^2\mathcal D\left[ a_j\right]\varrho_\alpha - \mathrm i \frac{1}{2}\left[\mathrm i(\alpha_j^\ast)^2 a_j^2 - \mathrm i\alpha_j^2 (\alpha_j^\dagger)^2,\varrho_\alpha\right] -\mathrm i \left|\alpha_j\right|^2\left[\mathrm i \alpha^\ast_j a_j - \mathrm i \alpha_j a_j^\dagger, \varrho_\alpha\right]\right\} + \mathcal O (a_j^3)
\end{aligned}
\end{equation}
\end{widetext}
We now realize that as long as condition (\ref{eq:VdPLattice}) i.e., the mean field equation, is satisfied, the linear terms vanish. If we neglect higher order terms, the quantum fluctuations are governed by master equation of Lindblad form:
\begin{equation}
\dot \varrho_\alpha = -\mathrm i \left[H_\alpha, \varrho_\alpha \right]+\sum\limits_j \left\{\kappa_1 \mathcal D\left[ a_j^\dagger\right]\varrho_\alpha+ 4 \kappa_2 \left|\alpha_j\right|^2\mathcal D\left[ \alpha_j\right]\varrho_\alpha\right\}
\end{equation}
with effective Hamiltonian 
\begin{equation}
H_\alpha = H_\mathrm S - \mathrm i \frac{\kappa_2}{2}\sum\limits_j \left(\alpha_j^2  a_j^\dagger  a_j^\dagger  - (\alpha_j^\ast)^2  a_j  a_j\right).
\end{equation}

\section{Specific form of the equation governing the covariance matrix}
\label{secApp:Covariance}

In this section we specify the entries of the matrices $\underline{\mathrm B}$ and $\underline{\mathrm D}$ governing the equation of motion of the covariance matrix $\underline{\mathrm C}$:
\begin{equation}
\dot{\underline{\mathrm C}} = \underline{\mathrm B}~\underline{\mathrm C}+ \underline{\mathrm C}~\underline{\mathrm B}^\intercal + \underline{\mathrm D}. 
\end{equation}
The entries of the matrices are given through Eq.~(\ref{eq:MasterEquationAlpha}). Specifically, $\underline{\mathrm B}$ is a block matrix, where the blocks on the diagonal $\underline{\mathrm B}_{jj}$ take the form
\begin{widetext}
\begin{equation}
\label{eq:BMatrix}
\underline{\mathrm B}_{jj} =\frac{1}{2} \left(\begin{matrix}
\kappa_1 - 4 \kappa_2 \left|\alpha_j\right|^2 - \kappa_2 [(\alpha_j^\ast)^2 +\alpha_j^2] & -2 \omega - 
 \mathrm i \kappa_2 [(\alpha_j^\ast)^2 - \alpha_j^2] \\
+2 \omega - 
 \mathrm i \kappa_2 [(\alpha_j^\ast)^2 - \alpha_j^2] & \kappa_1 - 4 \kappa_2 \left|\alpha_j\right|^2 + \kappa_2 [(\alpha_j^\ast)^2 +\alpha_j^2]
\end{matrix}\right)
\end{equation}
\end{widetext}
and the off-diagonal block matrices $\underline{\mathrm B}_{jk}$ for $j\neq k$ the form
\begin{equation}
\underline{\mathrm B}_{jj'} = \lambda_{jj'} \left(\begin{matrix}
0 & 1 \\
-1 & 0 
\end{matrix}\right).
\end{equation}
Furthermore, $\underline{\mathrm D}$ is a diagonal block matrix with entries $\underline{\mathrm D}_{jj}$ on the diagonal given by 
\begin{equation}
\label{eq:DMatrix}
\underline{\mathrm D}_{jj} = \frac{1}{2}(\kappa_1 + 4 \kappa_2 \left|\alpha_j\right|^2) \left(\begin{matrix}
1 & 0 \\
0 & 1
\end{matrix}\right).
\end{equation}

\section{\new{Comparison of the full dynamics with the effective model}}
\label{secApp:Validity}

\new{In this section we provide a comparison of the full quantum dynamics given by  Eqs.~(\ref{eq:GeneralHamiltonian}) and (\ref{eq:QauntumModel}), and the effective model described by Eqs.~(\ref{eq:MasterEquationAlpha}) and (\ref{eq:EffectiveHamiltonian}) for a single vdP oscillator and two coupled vdP oscillators in terms of Wigner representations and the synchronization measure defined in Eq.~(\ref{eq:AvgCompleteSynch}).}

\new{\subsection{Single vdP oscillator}}

\begin{figure*}
\begin{center}
\includegraphics[width=\linewidth]{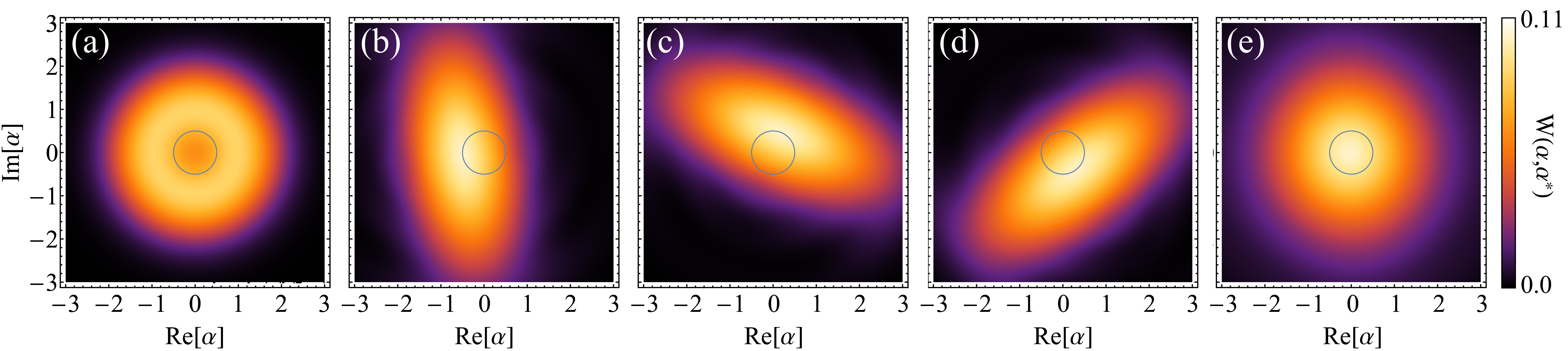}
\end{center}
\caption{Comparison of the exact model given by Eqs.~(\ref{eq:GeneralHamiltonian}) and (\ref{eq:QauntumModel}), and the effective model described by Eqs.~(\ref{eq:VdPLattice}), (\ref{eq:MasterEquationAlpha}) and (\ref{eq:EffectiveHamiltonian}) for a single vdP oscillator in terms of the Wigner function $W(\alpha,\alpha^\ast)$. (a) Without approximations the Wigner distribution at steady state has a donut shape centered at the origin. The corresponding limit cycle of the mean field solution is shown as blue solid line. (b)--(d) Wigner distribution of the effective model at different times (after a transient relaxation time), which is displaced by the mean field amplitude. (e) Time averaged Wigner distribution of the effective model together with the corresponding mean field limit cycle. Parameters: $\kappa_1 = 5\cdot 10^{-3} \omega_0$, $\kappa_2 = 2\kappa_1$.}
\label{fig:APPNew1}
\end{figure*}

\new{For a single vdP oscillator the dynamics is given by 
\begin{equation}
\label{eq:AppFullLindblad}
\dot{\varrho} = -\frac{\mathrm i }{\hbar}\omega_0 \left[a^\dagger a, \varrho\right] + \kappa_1 \mathcal D\left[a^\dagger\right]\varrho + \kappa_2 \mathcal D\left[a^2\right]\varrho.
\end{equation}
Its steady state is defined via $\dot\varrho_\mathrm{ss} =0$ and can be easily found numerically. The Wigner function $W(\alpha,\alpha^\ast)$  is a convenient way to visualize and compare the full quantum dynamics with the effective model derived in App.~\ref{secApp:EffectiveQM} \cite{Lee2013}. It represents a quasi-probability distribution in the space of coherent states. In Fig.~\ref{fig:APPNew1}(a) we show the Wigner function corresponding to the steady state of a single vdP oscillator defined via Eq.~(\ref{eq:AppFullLindblad}), which displays the typical donut-like shape expected for self-oscillating (quantum) systems. We also show the mean field limit cycle with radius $\bar A = \sqrt{\kappa_1/2\kappa_2}$ [cf.~\ref{subsec:MeanField}], which is however smaller than the maximum of the distribution. 
}

\new{The effective model is given by the Lindblad master equation
\begin{equation}
\label{eq:AppEffective}
\begin{aligned}
\dot \varrho_\alpha(t) =& -\frac{\mathrm i}{\hbar} \left[H_\alpha(t), \varrho_\alpha(t) \right] \\
&+\kappa_1 \mathcal D\left[ a^\dagger\right]\varrho_\alpha(t)+ 4 \kappa_2 \left|\alpha(t)\right|^2\mathcal D\left[ a\right]\varrho_\alpha(t)
\end{aligned}
\end{equation}
with time-dependent Hamiltonian 
\begin{equation}
H_\alpha(t) = \hbar\omega_0 a^\dagger a  - \mathrm i\hbar\frac{\kappa_2}{2}\left\{\left[\alpha(t)\right]^2  (a^\dagger)^2 - \left[\alpha^\ast(t)\right]^2  a^2\right\}.
\end{equation}
As we are interested in the (periodic) steady state dynamics, we may replace the mean field amplitude $\alpha(t)$ by its respective steady state dynamics $\bar\alpha(t) = \bar A \exp(-\mathrm i \omega_0 t)$, where we have chosen the arbitrary phase $\phi$ equal to zero (cf. Sec.~\ref{subsec:MeanField}). Furthermore, we choose the initial state $\varrho_\alpha(t=0) = \ket{0}\bra{0}$. As the time-dependent mean field amplitude $\bar\alpha(t)$ is periodic with period $T=2\pi/\omega_0$, the density matrix $\varrho_\alpha(t)$ will eventually also become periodic with the same periodicity. In Figs.~\ref{fig:APPNew1}(b)--(d) we show the Wigner function $W(\alpha,\alpha^\ast,t)$ at different times (after some short transient relaxation) and in Fig.~\ref{fig:APPNew1}(e) its time average. While at each time of the period, the Wigner function is squeezed in different directions and displaced by the mean field amplitude (the limit cycle is shown as blue solid line), the time-averaged Wigner function is a symmetric Gaussian distribution centered at the origin. Thus, within the effective model the donut shape of the Wigner distribution of a single vdP oscillator may not be observed because the mean field amplitude is smaller and the quantum fluctuations larger compared to the full dynamics without approximations. However, in the next section we will show that the effective model is still capable of capturing the signatures of synchronization of two coupled vdP oscillators. }

\new{\subsection{Two coupled vdP oscillators}}

\begin{figure}
\includegraphics[width=\columnwidth]{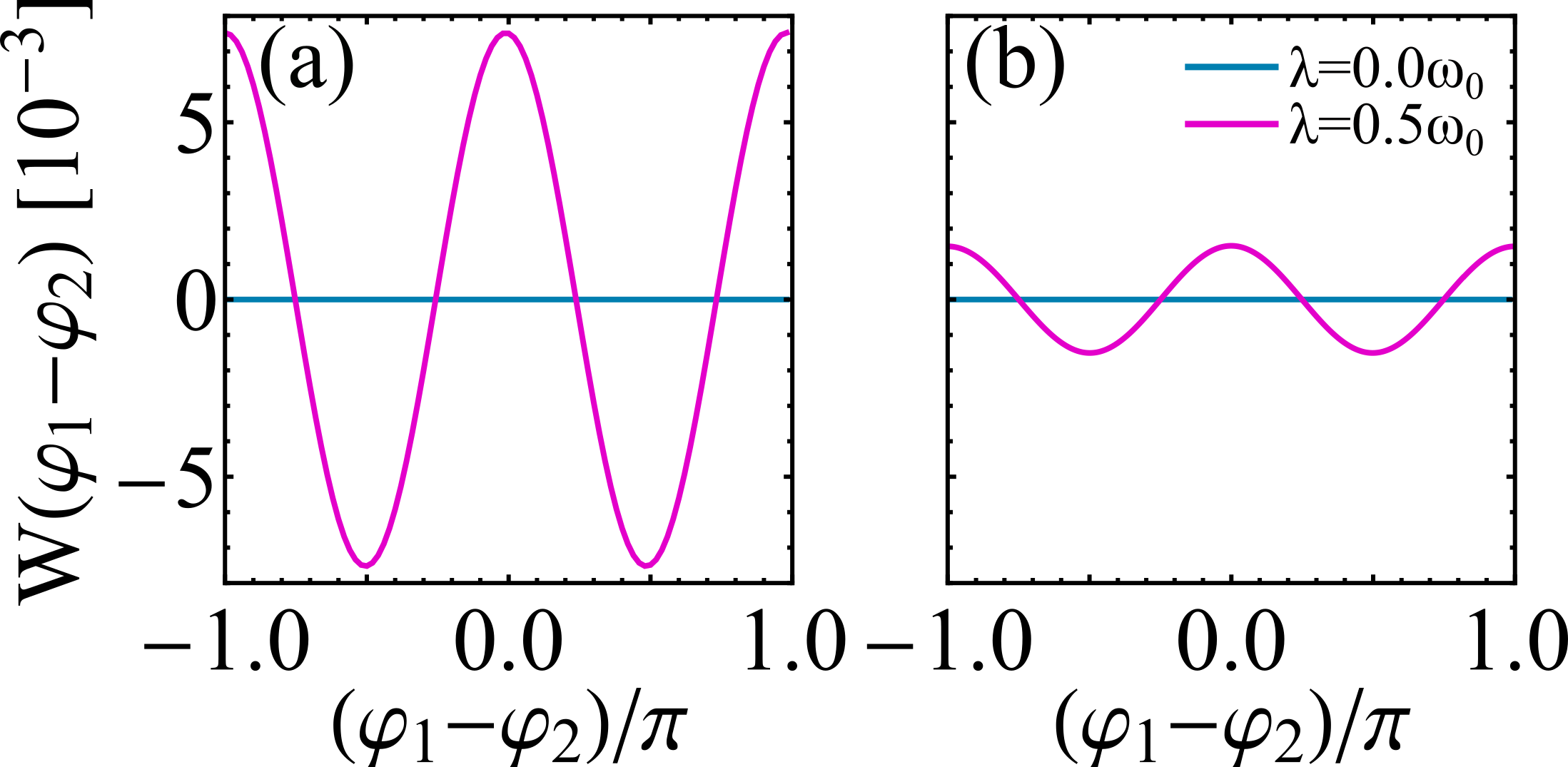}
\caption{Wigner distribution of the phase differences $W(\varphi_1 - \varphi_2)$ for two coupled vdP oscillators as a function of the phase difference with interaction strengths $\lambda = 0.0\omega_0$ (blue solid) and $\lambda = 0.5 \omega_0$ (pink solid) for (a) the exact model and (b) the effective model. Parameters: $\kappa_1 = 5\cdot 10^{-3} \omega_0$, $\kappa_2 = 2\kappa_1$.}
\label{fig:APPNew3}
\end{figure}

\new{The full quantum model of two coupled vdP oscillators is given by Eq.~(\ref{eq:QauntumModel}) with $j=1,2$ and system Hamiltonian
\begin{equation}
H = \hbar\sum\limits_j \omega_0 a^\dagger_j a + \hbar\lambda (a_1^\dagger a_2 + a_2^\dagger a_1).
\end{equation}
Following Ref.~\cite{Lee2013}, we characterize the two oscillator system in terms of the two mode Wigner function $W(\alpha_1,\alpha_1^\ast, \alpha_2,\alpha_2^\ast)$ and integrate $|\alpha_1|$, $|\alpha_2|$ and $\varphi_1+\varphi_2$, such that the Wigner function is a function of the phase difference $\varphi_1 - \varphi_2$ alone. In Fig.~\ref{fig:APPNew3} we show the Wigner function $W(\varphi_1-\varphi_2)-\pi/2$ as function of the phase difference between the two oscillators without coupling (blue) and with coupling $\lambda = 0.5\omega_0$ (pink). We choose the latter value as it is the largest coupling we consider in the main text. While the Wigner function is completely flat in the absence of interactions, it shows two peaks at $\varphi_1 - \varphi_2=0,\pi$ corresponding to in-phase and anti-phase synchronization respectively. Due to the strong quantum noise in the regime considered in this work, the peaks are quite small, which has also been reported in Ref.~\cite{Lee2013}.}

\begin{figure}
\includegraphics[width=0.8\columnwidth]{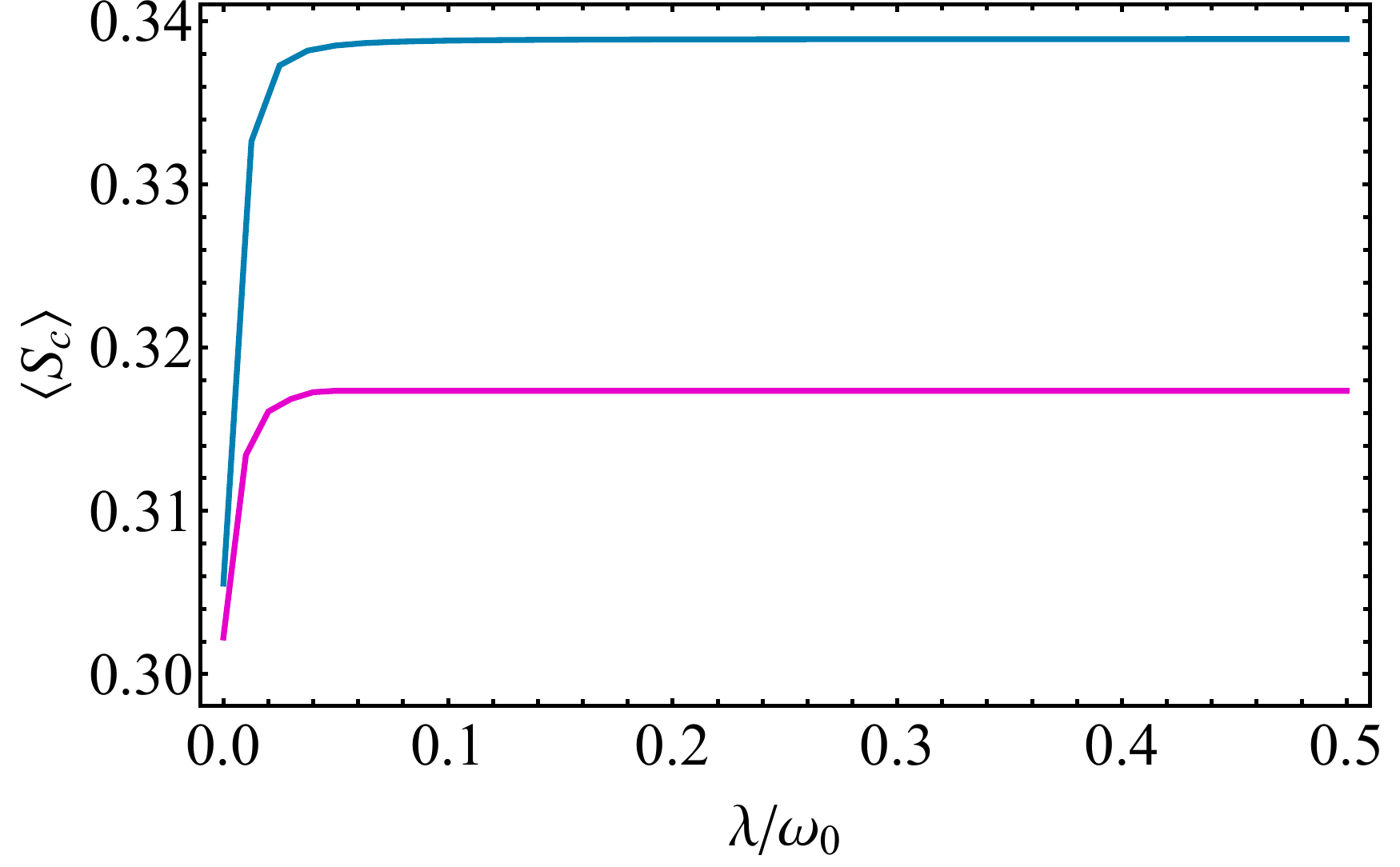}
\caption{Quantum synchronization measure $\left<S_\mathrm c\right>$ as a function of the coupling strength $\lambda$ between two coupled vdP oscillators. The blue line corresponds to the full dynamics of the exact model, whereas the pink line corresponds to the effective model. In both cases we observe an increase of the measure as the interaction is turned on, yet for the effective model, the maximum value of $\left<S_\mathrm c\right>$ remains always smaller compared to the exact model. Parameters: $\kappa_1 = 5\cdot 10^{-3} \omega_0$, $\kappa_2 = 2\kappa_1$.}
\label{fig:APPNew4}
\end{figure}

 \begin{figure*}
\begin{center}
\includegraphics[width=\linewidth]{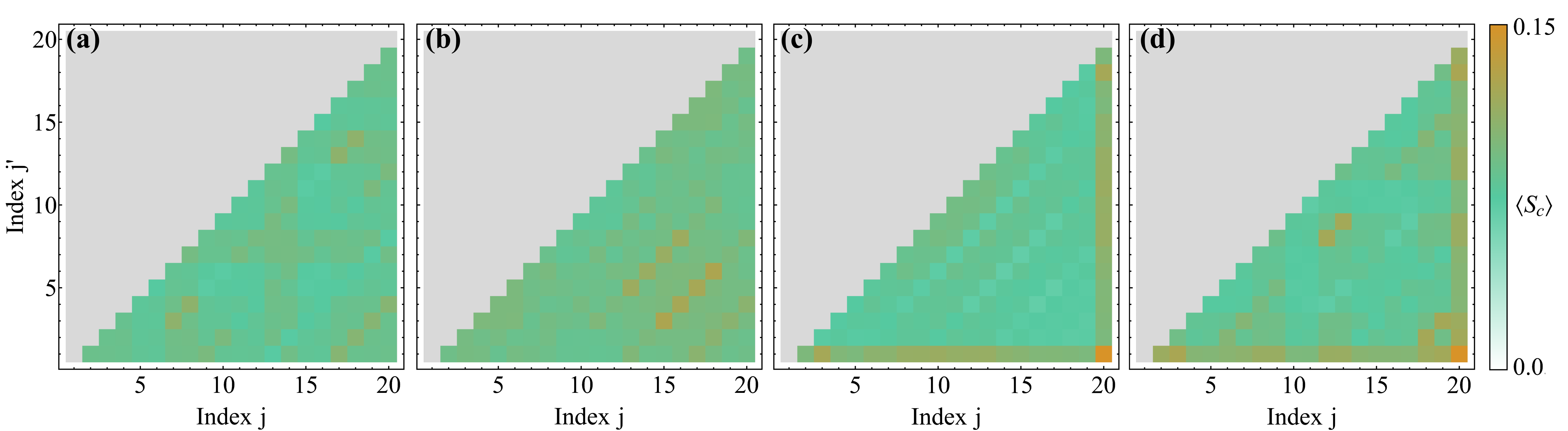}
\end{center}
\caption{Time-averaged quantum complete synchronization measure $\left<S_\mathrm c\right>$ between lattice site $j$ and $j'$ of the vdP SSH network corresponding to the mean field dynamics shown in Fig.~2, i.e., with an eigenstate $\boldsymbol \alpha^{(l)}$ of the topological Hamiltonian $H_\mathrm{SSH}$ [cf. Eq.~(16)] as initial conditions and for different coupling strength (a) $\delta \lambda=-0.8 \lambda_0$, (b) $\delta \lambda = - 0.4\lambda_0$, (c) $\delta \lambda = 0.4 \lambda_0$, and (d) $\delta \lambda = 0.6 \lambda_0$. With an eigenstate as initial state, additional structures emerge compared to random initial conditions. Nevertheless, $\left<S_\mathrm c\right>$ between the edges ($j'=1$ and $j=20$) is still the largest quantity. Parameters: $\kappa_1 = 5\cdot 10^{-3} \omega_0$, $\kappa_2 = 2\kappa_1$, $\lambda_2 = 0.25 \omega_0$, $\omega_0 t_i = 2 \cdot 10^4$, $\omega_0 t_f=2.3\cdot 10^4$.}
\label{fig:APP1}
\end{figure*}

\new{For two coupled vdP oscillators the effective dynamics are given by Eq.~(\ref{eq:EffectiveHamiltonian}) with effective Hamiltonian
\begin{equation}
\begin{aligned}
H_\alpha(t) =& \hbar\sum\limits_j \omega_0 a^\dagger_j a + \hbar\lambda (a_1^\dagger a_2 + a_2^\dagger a_1) \\
&- \mathrm i\hbar\frac{\kappa_2}{2}\sum\limits_j \left\{\left[\alpha_j(t)\right]^2  (a_j^\dagger)^2 - \left[\alpha_j^\ast(t)\right]^2  a_j^2\right\},
\end{aligned}
\end{equation}
where the mean field amplitudes $\alpha_j(t)$ follow the dynamics 
\begin{equation}
\begin{aligned}
\dot \alpha_1 &= -\mathrm i \omega_0 \alpha_1 -\mathrm i \lambda \alpha_2 +\frac{\kappa_1}{2}\alpha_1 -\kappa_2 |\alpha_1|^2\alpha_1, \\
\dot \alpha_2 &= -\mathrm i \omega_0 \alpha_2 -\mathrm i \lambda \alpha_1 +\frac{\kappa_1}{2}\alpha_2 -\kappa_2 |\alpha_1|^2\alpha_2.
\end{aligned}
\end{equation}
In the case without interaction ($\lambda =0$) the mean field amplitude of each oscillator will eventually reach the steady state $\bar\alpha_j(t) = \bar A \exp(-\mathrm i \omega_0 t+ \varphi_j)$ with an arbitrary phase difference $\varphi_1 - \varphi_2$ set by the initial conditions. In the case of large interaction strengths the two oscillators will synchronize, however, either in-phase or anti-phase, again set by the initial conditions. It is thus necessary to average over the initial conditions in order to properly compare it to the full quantum dynamics. As discussed previously, the initial condition of the effective density matrix is given by $\varrho_\alpha(t=0) = \bigotimes_j \ket{0_j}\bra{0_j}$. In Fig.~\ref{fig:APPNew3}(b) we show the Wigner function $W(\varphi_1-\varphi_2)-\pi/2$ as function of the phase difference between the two oscillators for $\lambda = 0$ (blue) and $\lambda = 0.5\omega_0$ (pink). Similar to the full quantum model shown in  Fig.~\ref{fig:APPNew3}(a), we observe that if the interaction is turned off the Wigner function is completely flat, while it is peaked at $\varphi_1 - \varphi_2=0,\pi$ with the coupling turned on, however, compared to panel (a) the peaks are smaller. The reason is that the effective model overestimates the quantum fluctuations leading to less pronounced synchronization. This is consistent with the observations of Ref.~\cite{Lee2013}, where for $\kappa_2 = 10\kappa_1$ the height of the peaks is in the same order of magnitude.}

\new{In the main manuscript we quantify the level of synchronization between two vdP oscillators in terms of the synchronization measure defined in Eq.~(\ref{eq:AvgCompleteSynch}) rather than the Wigner distribution. We therefore also compare this quantity for the full quantum model and the effective model. In Fig.~\ref{fig:APPNew4} we show $\left<S_\mathrm c\right>$  as a function of the coupling strength $\lambda$ for two coupled vdP oscillators without (blue) and with approximations (pink). In both cases, $\left<S_\mathrm c\right>$ is finite even though the oscillators do not interact, which is due to the fact that both oscillators are equivalent. As the interaction increases, also $\left<S_\mathrm c\right>$ increases until it quickly approaches a plateau. In accordance with the previous observations regarding the Wigner function, also the synchronization measure is overall smaller for the effective model than for the full dynamics.}

\section{Quantum signatures in the SSH model for eigenstates as initial conditions}
\label{secApp:QuantumFlucEV}

In this section we provide for completeness the quantum synchronization measure $\left<S_\mathrm c\right>$ corresponding to the mean field  amplitudes shown in Fig.~\ref{fig:Fig2}, that is we choose eigenstates $\boldsymbol \alpha^{(l)}$ of the topological Hamiltonian $H_\mathrm{SSH}$ [cf. Eq.~(\ref{eq:SSHHamiltonian})] as initial states (cf. Sec.~\ref{subsec:TopologySSH}). In Figs.~\ref{fig:APP1}(a)-(d) we show the time-averaged quantity $\left<S_\mathrm c\right>$ between lattice site $j$ and $j'$ of the vdP chain for different coupling strengths (a) $\delta \lambda=-0.8 \lambda_0$, (b) $\delta \lambda = - 0.4\lambda_0$, (c) $\delta \lambda = 0.4 \lambda_0$, and (d) $\delta \lambda = 0.6 \lambda_0$. In the trivial phase we observe complete synchronization of all vdP oscillators for the mean fields shown in Figs.~\ref{fig:Fig2}(b) and (c) with oscillation amplitudes determined by the amplitudes of the initial state. Moreover, the amplitudes are symmetric with symmetry center located in the middle of the chain, i.e. between oscillator $j=10$ and $j=11$. For the quantum model, this symmetry is reflected in the synchronization measure shown in Figs.~\ref{fig:APP1}(a) and (b), i.e., $\left<S_\mathrm c(j',j)\right> = \left<S_\mathrm c(j'=N+1-j,j=N+1-j')\right>$, for example $\left<S_\mathrm c(j'=3,j=7)\right> = \left<S_\mathrm c(j'=14,j=18)\right>$ in Fig.~\ref{fig:APP1}(a). However, besides this observation, comparing the mean field dynamics with the emergent quantum synchronization remains challenging and will be investigated further in future work. 

The bulk oscillators in the topological phase shown in Figs.~\ref{fig:APP1}(b) and (c) corresponding to the mean field amplitudes of Figs.~\ref{fig:Fig2}(d) and (e), respectively, exhibit similar behavior as discussed previously; notably $\left<S_\mathrm c(j',j)\right>$ reflects the symmetry of the system. However, in addition the oscillators located at the two edges of the chain ($j'=1$ and $j=20$) exhibit the largest value of the quantum synchronization measure in the vdP chain. It is worth mentioning that the oscillators located at the boundaries live on different sublattices ($j'=1$ is odd and $j=20$ is even) and furthermore exhibit in-phase synchronization of the mean field amplitudes; see Figs.~\ref{fig:Fig2}(d) and (e). 

\section{Including an additional linear loss channel}
\label{secApp:LinearLoss}

For the experimental realization of a vdP oscillator with optomechanics as discussed in Sec.~\ref{sec:Expoeriments}, there exists an additional linear one-phonon loss channel. However, we will show in this section that such an additional dissipative term does not affect the results discussed in the main text as long as it is small compared to the linear gain channel. Including this additional loss term, the Lindblad master equation is given by
\new{\begin{equation}
\dot{\varrho} = -\frac{\mathrm i}{\hbar}\left[H,\varrho\right] +  \sum\limits_{j} \left\{\kappa_1\mathcal D\left[a_j^\dagger\right]\varrho +\kappa_2\mathcal D\left[a_j^2\right]\varrho + \bar\gamma\mathcal D\left[a_j\right]\varrho \right\} .
\end{equation}}
Within in the mean field approximation, the term proportional to $\bar\gamma$ only appears as a shift of the dissipation rate $\kappa_1$, i.e.,  with the definition $\tilde\kappa_1 = \kappa_1- \bar\gamma$ the dynamics is fully equivalent to Eq.(\ref{eq:MasterEquationAlpha}) assuming that $\bar\gamma \ll \kappa_1$. The latter condition must be fulfilled in order to counteract the nonlinear damping ($\kappa_2$) and not simply decay into the ground state. 

In terms of the quantum fluctuations about these mean field amplitudes, the additional $\bar\gamma$ term appears in the matrices $\underline{\mathrm B}$ and $\underline{\mathrm D}$ governing the dynamics of the covariance matrix; see Eq.~(\ref{eq:DiffEqCorrelation}). Similar to the mean field approximation, in $\underline{\mathrm B}$ this will only result in an effective dissipation rate $\kappa_1\to \tilde\kappa_1= \kappa_1-\bar\gamma$ in Eq.~(\ref{eq:BMatrix}). However, the sign is flipped for the effective dissipation rate in $\underline{\mathrm D}$ [cf.~Eq.~(\ref{eq:DMatrix})], which is then given by
\begin{equation}
\underline{\mathrm D}_{jj} = \frac{1}{2}(\kappa_1 + \bar\gamma + 4 \kappa_2 \left|\alpha_j\right|^2) \left(\begin{matrix}
1 & 0 \\
0 & 1
\end{matrix}\right).
\end{equation}
We have checked numerically that for $\bar\gamma = 0.1\kappa_1$ there are no noticeable changes in the dynamics or synchronization behavior, such that we are confident that in fact a linear dissipation channel does not alter the discussed results.

\end{document}